\documentclass[11pt,a4paper]{article}

\usepackage{graphicx}
\usepackage{natbib}
\usepackage{amsmath,amssymb}
\usepackage{setspace}
\usepackage{hyperref}

\normalsize

\title{Designs for Generalized Linear Models}
\author{Anthony C. Atkinson\footnote{Email: {\tt a.c.atkinson@lse.ac.uk}}
 \and  David C. Woods\footnote{Email: {\tt
d.woods@southampton.ac.uk}}\\
$^\star$London School of Economics and Political Science, UK \and $^\dagger$University of Southampton, UK
}
\date{}

\begin{document}
\maketitle
\newpage
\tableofcontents
\newpage
\pagenumbering{arabic}


\section{Introduction}
\label{introsec}

The methods of experimental design described in the majority of the earlier chapters  are appropriate if the continuous response, perhaps after transformation, has independent errors with a variance that is known up to a  multiplicative constant.  (An exception is Chapter~5, which describes designs for correlated errors). However, this is not a characteristic of the Poisson and binomial
distributions, where there is a strong relationship for these discrete random variables between mean and variance. The main emphasis of this chapter is on designs for generalized linear models appropriate for data from these and other distributions.

The classic account of generalized linear models is \citet{mcc+n:89}. Issues in the design of experiments for these models are reviewed by \citet{Khuri+:2007}; in addition to the methods of optimal experimental design, they consider stochastic approximation \citep{robb+monro:51} and adaptations of response surface methodology (\citealp{b+d:63}, and Chapter~10 of this Handbook). Their emphasis is mainly on models with a single explanatory variable. On the other hand, the review of \citet{aca:2005c} focuses on optimal design and models with several explanatory variables as, to some extent,  does the more recent review of \citet{StuffGLM;2012}, where the emphasis is towards analytical results. Here we follow the approach of \citeauthor{aca:2005c}, but focus on more recent results and on computational methods.

The assumptions of normality and constancy of variance for regression models enter the criteria of optimal design through the form of the information matrix $\boldsymbol{X}^{\prime}\boldsymbol{X}$, where, as in other chapters, $\boldsymbol{X}$ is the $n \times p$ model, or extended design, matrix. Other forms of information matrix arise from other distributions. See  \citet{aca+val+:2014}. Given the appropriate information matrix, the principles of optimal design are the same as those described in earlier chapters. In designs for generalized linear models, the asymptotic covariance matrix of the parameters of the linear model is of the form $\boldsymbol{X}^{\prime}\boldsymbol{U}\boldsymbol{X}$, where the $n \times n$ diagonal matrix of weights $\boldsymbol{U}$ depends on the parameters of the linear predictor, on the error distribution and on the link between them. The dependence of the designs on parameters whose values are unknown prior to experimentation means that, in general, designs for generalized linear models require some specification of prior information.

In \S\ref{glmsec}, we briefly review the class of generalized linear models, with particular emphasis on models for binomial, Poisson and gamma data. Some fundamental ideas in optimal design of experiments are reviewed in \S\ref{optdessec}, and the optimality criteria employed in this chapter are introduced. We emphasize the reliance of optimal designs on the unknown parameter values, and discuss methods of overcoming this dependence. Locally optimal designs are introduced in \S\ref{localsec} for logistic models for binomial
data with a single explanatory variable. In \S\ref{localsevfacsec}  we move on to  designs for binomial, Poisson and gamma data with several explanatory variables. This latter section also includes results on Bayesian designs. In \S\ref{corr}, we discuss designs for dependent non-normal data, for example, arising from blocked experiments, and demonstrate optimal design for generalized linear mixed models through an example. In \S\ref{extensec}, we give some extensions and suggestions for further reading.

\section{Generalized linear models}
\label{glmsec}

\subsection{The family of models}
\label{famsec}

The family of generalized linear models extends normal theory regression to any distribution belonging to the one-parameter exponential family. As well as the normal (with known variance), this includes the gamma, Poisson and binomial distributions, all of which are important in the analysis of data. The three components of a generalized linear model are:

\begin{enumerate}
\item A distribution for the univariate response $y$ with mean $\mu$.
\item A linear predictor $\eta = \boldsymbol{f}^\prime(\boldsymbol{x})\boldsymbol{\theta}$ where $\boldsymbol{f}(\boldsymbol{x})$ is a $p$-vector of known functions of the $k$ explanatory variables $\boldsymbol{x}$, and $\boldsymbol{\theta}$ is a $p$-vector of unknown model parameters.
\item A link function $g(\mu) = \eta$, relating $\boldsymbol{x}$ to the mean $\mu$.
\end{enumerate}

The distribution of $y$  determines the relationship between the mean and the variance of the observations. The variance is of the form

\begin{equation}
\mbox{var}(y) = \phi V(\mu)\,, \label{varfn}
\end{equation}

\noindent where $\phi$ is a `dispersion parameter', equal to $\sigma^2$ for the normal distribution and equal to one for the binomial and Poisson distributions. The variance function $V(\mu)$ is specific to the error distribution.

The information matrix for a single observation at a point $\boldsymbol{x}$ is

\begin{equation}
\boldsymbol{M}(\boldsymbol{x};\,\boldsymbol{\theta}) = u(\boldsymbol{x})\boldsymbol{f}(\boldsymbol{x})\boldsymbol{f}^\prime(\boldsymbol{x})\,,
\label{eleminfweights}
\end{equation}

\noindent with the weights for individual observations  given by

\begin{equation}
u(\boldsymbol{x}) = V^{-1}(\mu)\left(\frac{d\mu}{d\eta}\right)^2. \label{eq6}
\end{equation}

\noindent These weights depend both on the distribution of $y$ and on the link function.

\subsection{The normal distribution}
\label{normalsec}

The linear multiple regression model  can be written as

\begin{equation}
\mbox{\rm E}(y) = \mu = \eta = \boldsymbol{f}^\prime(\boldsymbol{x})\boldsymbol{\theta}\,,
\label{eq4}
\end{equation}

\noindent where $\mu$, the mean of $y$ for given $\boldsymbol{x}$, is equal to the  linear predictor $\eta$. In~\eqref{varfn}, $V(\mu) = u(\boldsymbol{x}) = 1$ and, in this simple case, $\phi = \sigma^2$.

Important design problems arise with extensions to this model, particularly those in which the variance is parameterized through a link function and linear predictor, that may include parameters in common with the linear predictor for the mean (\citealp{muir:82,jan+:88,a+cook:94} and \citealp[\S6.3.1]{valerii+sergei:2014}). Some  references for designs related to the extension of the model to include transformation of the response and to transformation of both sides of the model are in \S\ref{extensec}.

\subsection{Binomial data}
\label{binsec}

For the binomial distribution, the variance function~\eqref{varfn} is

\begin{equation}
V(\mu) = \mu(1-\mu)\,.
\label{binv}
\end{equation}

\noindent In models for binomial data with $R$ successes in $n$ observations, the response $y$ is defined to be $R/n$. The link function should be such that, however the values of $\boldsymbol{x}$ and $\boldsymbol{\theta}$ vary, the mean $\mu$ satisfies the physically meaningful constraint that $0 \le \mu \le 1$. We list four link functions that have been found useful in the analysis of data.

\begin{enumerate}
\item {\bf Logistic}:
\begin{equation} \eta = \log \left(
\frac{\mu}{1- \mu} \right)\,.
\label{logisticl}
\end{equation}

\noindent The ratio $\mu/(1- \mu)$ is the odds that $y = 1$. In the logistic
model the log odds is therefore equal to the linear predictor.
Apart from a change of sign, the model is unaltered if ``success"
is replaced with ``failure". For this canonical link, it follows from calculation of $d\eta/d \mu$ that

\begin{equation}
u(\boldsymbol{x}) = \mu(1-\mu)\,,
\label{logiwt}
\end{equation}

\noindent a simpler form than for the other three link functions we shall discuss.

\item {\bf Probit}:
\begin{equation} \eta = \Phi^{-1}(\mu)\,,
\label{probitl}
\end{equation}

\noindent where $\Phi$ is the normal cumulative distribution function. This
link has very similar properties to those of the logistic link. In this case

\begin{equation}
u(\boldsymbol{x}) = \frac{\phi^2(\eta)}{\Phi(\eta)\{1-\Phi(\eta)\}}\,.
\label{probiwt}
\end{equation}

\noindent Here $\phi$ is the standard normal density.

\item {\bf Complementary log-log}:
\begin{equation} \eta = \log \{ - \log (1 - \mu)\},
 \label{cll}
\end{equation}

\noindent for which

\begin{equation}
u(\boldsymbol{x}) = \frac{(1-\mu)}{\mu}\{\log(1-\mu)\}^2\,.
\label{cllogwt}
\end{equation}

\noindent  The complementary log-log link is not symmetric in success and failure, so providing a model with properties distinct from those of the logistic and probit links. Interchanging success and failure gives the following log-log link.

\item {\bf Log-log}:
\begin{equation} \eta = \log ( - \log  \mu)\,.
\label{ll}
\end{equation}
\end{enumerate}

\noindent A plot of these four link functions is in Figure~4.1 of \citet{mcc+n:89}. \citet[\S6.18]{a+r:2000}  describe a fifth link, the arcsine link, which has some desirable robustness properties for binary data. In our examples we only calculate designs for the logistic link. \citet[\S22.4.3]{ADT:2007} compare designs for logistic and complementary log-log links when there is a single explanatory variable.

\subsection{Poisson data}
\label{poisssec}

For Poisson data, where $V(\mu) = \mu$,  we require that $\mu > 0$. The log link, $\log \mu = \eta$, is standard for the analysis of Poisson data in areas such as medicine, social science and economics; see, for example, \citet[][ch. 9]{agr:2002}, \citet{winkelmann}, and \citet{emm2011}. This link leads to models with $\mu = \exp \eta$, which satisfy the constraint on values of $\mu$, and weights $u(\boldsymbol{x}) = \mu$.

\subsection{Gamma data}
\label{gamsec}

The gamma family is one in which the correct link is often in doubt. The physical requirement is again that $\mu > 0$. A useful, flexible family of links  that obeys this constraint is the Box and Cox family, in which

\begin{equation}
g(\mu) =  \left\{\begin{array}{lr}(\mu^{\lambda} - 1)/\lambda &
(\lambda \ne 0)\; \\ \log \mu & (\lambda = 0)\,. \end{array} \right.
\label{eqg1}
\end{equation}

\noindent See \citet{b+c:64} for the use of this function in the transformation of data.

Differentiation of \eqref{eqg1} combined with \eqref{varfn} shows that the weight for the gamma distribution with this link
family is

\begin{equation}
u(\boldsymbol{x}) = \mu^{-2\lambda}\,.
\label{eqg3}
\end{equation}

\noindent When $\lambda = 0$, that is for the log link, the weights in~\eqref{eqg3} are equal to one. It therefore follows that optimal designs for gamma models with this link and $\lambda = 0$  are identical to optimal designs for regression models with the same linear predictors. Unlike designs for binomial and Poisson generalized linear models,  the designs when $\lambda = 0$ do not depend on the parameter $\boldsymbol{\theta}$.

This link is seemingly equivalent to the power family of links

\begin{equation}
g(\mu) = \mu^{\kappa}\,,
\label{powerlink}
\end{equation}

\noindent for which differentiation shows that

\begin{equation}
u(\boldsymbol{x}) = \dfrac{1}{\kappa^2}\dfrac{1}{\mu^{2\kappa}} = \dfrac{1}{\kappa^2}\dfrac{1}{\eta^2}\,.
\label{powerlinku}
\end{equation}

\noindent Since $1/\kappa^2$ is constant over ${\cal X}$, the optimal design depends on   $\eta$, but not on $\kappa$. The situation is similar to that for normal theory regression, in which D-optimal designs are independent of the value of the variance $\sigma^2$, although the information matrix is a function of that parameter. In \S\ref{twolinkssec} we discuss  the relationship of the designs produced by these two links.

The gamma model is often an alternative to response transformation (\S\ref{extensec}). In particular, with a log link, it  may be hard to distinguish the gamma from a linear regression model with logged response. A discussion is in \S\S8.1 and 8.3.4 of \citet{mcc+n:89} with examples of data analyses in \S7.5 of  \citet{mmvr2010}.

\section{Optimal experimental designs}

\label{optdessec}

\subsection{Theory}

\label{theorsec}

The first exposition of optimal design in its modern form is \citet{kief:59}, although the subject goes back to \citet{smik:18} (see Chapter 1 , especially \S 1.9.3). Book length treatments include \citet{fed:72}, \citet{paz:86} and \citet{puke:93}. The focus of \citet{silv:80} and \citet{fed+hack:97} is on the mathematical theory; \citet*{ADT:2007} and \citet{wong+berger:2009} are more concerned with applications, whereas \citet{Pete+Brad:2011} introduce theory and examples via the JMP software. \citet{pronz+paz:2013} present the theory of optimal design for nonlinear models, while  \citet{valerii+sergei:2014} illustrate their theory with pharmaceutical applications, particularly dose finding.

As in \S3.1.2 of Chapter~2, where interest was in linear models, we let an experimental design $\xi$ place a fraction $w_i$ of the experimental trials at the conditions $\boldsymbol{x_i}$. A design with $t$ points of support is written as

\begin{equation}
  \xi = \left\{ \begin{array}{cc}
 \boldsymbol{x_1} & \boldsymbol{x_2} \ldots \boldsymbol{x_t} \\
  w_1 & w_2 \ldots w_t
  \end{array} \right\},
\label{optdes}
\end{equation}

\noindent where $w_i > 0$ and $\sum_{i=1}^t w_i= 1$. There are thus two sets of weights: $w_i$ which determine the proportion of experimentation at $\boldsymbol{x_i}$ and $u(\boldsymbol{x_i})$ which, from \eqref{eleminfweights}, partially determine the amount of information  from observations at that point. Any realisable experimental design for a total of $n$ trials will require that the weights are ratios of integers, that is $w_i = r_i/n$, where $r_i$ is the number of replicates at condition $\boldsymbol{x_i}$. Such designs are called \textit{exact} and are labeled $d_n$.  The mathematics of finding optimal experimental designs and demonstrating their properties is greatly simplified by consideration of \textit{continuous} designs in which the integer restriction is ignored.

The resulting design is a list of conditions under which observations are to be taken. The order in which the observations are made is usually also important; for example, the order should be randomized subject to any restrictions imposed by, for example, blocking factors (\S\ref{contblock}).

Optimal experimental designs minimize some measure of uncertainty of the parameter estimators, typically a function of the
information matrix. They require the specification of

\begin{enumerate}
\item A model, or set of models, of interest. For generalized linear models the specifications will include a set of parameter values $\boldsymbol{\theta} \in \Theta$, perhaps with an accompanying prior distribution $p(\boldsymbol{\theta})$.

\item A design criterion; for example, the minimization of a function of one or more information matrices.

\item A design region $\mathcal{X}\subseteq \Re^k$ to which the $\boldsymbol{x}_i$ belong.
\end{enumerate}

The information matrix for the design $\xi$ with $t$ support points is, from~\eqref{eleminfweights},

\begin{equation}
\boldsymbol{M}(\xi;\,\boldsymbol{\theta}) = \sum_{i=1}^t w_i \boldsymbol{M}(\boldsymbol{x}_i;\,\boldsymbol{\theta}) = \sum_{i=1}^t w_i u(\boldsymbol{x}_i)\boldsymbol{f}(\boldsymbol{x}_i)\boldsymbol{f}^\prime(\boldsymbol{x}_i)\,.
\label{infomat}
\end{equation}

\noindent As we are concerned with generalized \textit{linear} models, the parameter values enter only through the GLM weights $u(\boldsymbol{x}_i)$.

For  continuous designs we consider minimization of the general measure of imprecision $\Psi\{\boldsymbol{M}(\xi;\,\boldsymbol{\theta})\}$. Under very mild assumptions, the most important of which are the compactness of $\mathcal{X}$ and the convexity and differentiability of $\Psi$,
designs that minimize $\Psi$ also satisfy a second criterion. The relationship between these two provides a `general equivalence theorem', one form of which was introduced by \citet{k+w:60}. See \S2.3 of this book for a discussion of such theorems for linear models.

Let the measure $\bar{\xi}$ put unit mass at the point $\boldsymbol{x}$ and let the measure $\xi^{\alpha}$ be given by

\begin{equation}
\xi^{\alpha} = (1- \alpha)\xi + \alpha \bar{\xi}\qquad(0 \le \alpha \le 1).
\label{dirmeas}
\end{equation}

\noindent Then, from~\eqref{infomat},

\begin{equation}
\boldsymbol{M}(\xi^{\alpha};\,\boldsymbol{\theta}) = (1- \alpha)\boldsymbol{M}(\xi;\,\boldsymbol{\theta}) + \alpha \boldsymbol{M}(\bar{\xi};\,\boldsymbol{\theta})\,.
\label{dirmat}
\end{equation}

\noindent Accordingly, the derivative of $\Psi$ in the direction $\bar{\xi}$ is

\begin{equation}
\psi(\boldsymbol{x},\xi;\,\boldsymbol{\theta}) = \underset{\alpha \rightarrow 0^+}{\lim}\,
\frac{1}{\alpha}[\Psi\{(1-\alpha)\boldsymbol{M}(\xi;\,\boldsymbol{\theta}) + \alpha \boldsymbol{M}(\bar{\xi};\,\boldsymbol{\theta})\} -
\Psi\{\boldsymbol{M}(\xi;\,\boldsymbol{\theta})\}]\,.
\label{psideriv}
\end{equation}

The values of $\boldsymbol{M}(\xi;\,\boldsymbol{\theta})$ and of the Fr\'{e}chet derivative $\psi(\boldsymbol{x},\xi;\,\boldsymbol{\theta})$ depend on the parameter value $\boldsymbol{\theta}$, as will the design $\xi^*$ minimizing $\Psi\{\boldsymbol{M}(\xi;\,\boldsymbol{\theta})\}$. Variation of $\xi^*$ with $\boldsymbol{\theta}$ is one of the themes of this chapter. In order to incorporate uncertainty in $\boldsymbol{\theta}$ we define (pseudo) Bayesian criteria and state the equivalence theorem in a form that incorporates the  prior distribution $p(\boldsymbol{\theta})$ through use of

\begin{equation}
 \Psi\{\xi\} = E _{\boldsymbol{\theta}}\Psi\{\boldsymbol{M}(\xi;\,\boldsymbol{\theta})\}
  = \int_{\Theta}\Psi\{\boldsymbol{M}(\xi;\,\boldsymbol{\theta})\}p(\boldsymbol{\theta})d\boldsymbol{\theta}\,,
\label{bayscrit}
\end{equation}


\noindent and

\begin{equation}
\psi(\boldsymbol{x},\xi) = {\rm E} _{\boldsymbol{\theta}}\psi(\boldsymbol{x},\xi;\,\boldsymbol{\theta})\,.
\label{bayscrit2}
\end{equation}

{\bf The General Equivalence Theorem} states the equivalence of
the following three conditions on $\xi^*$:
\begin{enumerate}

\item The design $\xi^*$ minimizes $\Psi(\xi)$.

\item The design $\xi^*$ maximizes the minimum over ${\mathcal X}$  of $\psi(\boldsymbol{x},\xi)$.

\item The minimum over $\mathcal{X}$ of $\psi(\boldsymbol{x},\xi^*)$ is equal to zero,  this minimum occurring at the
points of support of the design.

\end{enumerate}

\noindent As a consequence of 3, we obtain the further condition:

\begin{enumerate}
\setcounter{enumi}{3}
\item For any non-optimal design $\xi$ the  minimum over $\mathcal{X}$ of $\psi(\boldsymbol{x},\xi) < 0.$
\end{enumerate}

\noindent The proof of this theorem follows \citet{whit:73}. See \citet{chal+l:89} and \citet{wl2011} for general equivalence theorems developed for, and applied to, optimal designs for generalized linear models.

As we illustrate later, the theorem provides methods for the construction and checking of designs. However, it says nothing about $t$, the number of support points of the design. If $p(\boldsymbol{\theta})$ is a point prior, putting all mass at a single value $\boldsymbol{\theta}_0$, the designs are called \textit{locally optimal}.  A bound on $t$ can then be obtained from the nature of $\boldsymbol{M}(\xi;\,\boldsymbol{\theta}_0)$, which is a symmetric $p\times p$ matrix. Due to the additive nature of information matrices~(\ref{infomat}), it follows from Carath\'eodory's Theorem that the information matrix of a design can be represented as a weighted sum of, at most, $p(p+1)/2+1$ rank-one information matrices. The maximum number of support points is therefore $p(p+1)/2+1$. A careful discussion is given by \citet[\S8.3]{puke:93}, with a shorter treatment by \citet[\S2.4.1]{valerii+sergei:2014} who state a simple and usually satisfied condition under which the maximum number of support points reduces to $p(p+1)/2$. In the examples that follow the number of support points of optimal designs is  usually appreciably less than this; often as few as $p$ is optimal. Of course, such designs with minimum support  provide no means of model checking (see \S\ref{extensec}). For more general priors
$p(\boldsymbol{\theta})$, the number of support points may be larger, increasing with the variance of the prior. \citet[p.~300]{ADT:2007} give examples for one-parameter nonlinear normal models in which the optimal designs have up to five support points.

\subsection{D-optimal designs}
\label{doptsec}

The most widely used design criterion is that of D-optimality (see Chapter 2) in which $\Psi\left\{\boldsymbol{M}(\xi;\,\boldsymbol{\theta})\right\} = - \log |\boldsymbol{M}(\xi;\,\boldsymbol{\theta})|$, so that the log determinant of the information matrix is to be maximized. Then~\eqref{bayscrit} becomes

\begin{equation}
 \Psi(\xi) = - E _{\boldsymbol{\theta}}\log |\boldsymbol{M}(\xi;\,\boldsymbol{\theta})|\,.
\label{baysdcrit}
\end{equation}

\noindent This (pseudo) Bayesian D-optimality criterion has been used to find designs for generalized linear models by \citet{chal+l:89} and \citet{woods+:2006}, amongst others. See \S\ref{bayesuncsubsec} for a comment on the distinction between designs under such criteria and truly Bayesian designs.

\citet[p.~68 and \S10.1]{valerii+sergei:2014} provide the mathematical results, including matrix differentiation,  required for calculation of the Fr\'{e}chet derivative  \eqref{psideriv}. The expected derivative~\eqref{bayscrit2}   then becomes

\begin{eqnarray}
\psi(\boldsymbol{x},\xi)& = &  E _{\boldsymbol{\theta}}\left\{p - u(\boldsymbol{x})\boldsymbol{f}^\prime(\boldsymbol{x})\boldsymbol{M}^{-1}(\xi;\,\boldsymbol{\theta})\boldsymbol{f}(\boldsymbol{x})\right\} \nonumber \\
& = & p - E _{\boldsymbol{\theta}}\left\{u(\boldsymbol{x})\boldsymbol{f}^\prime(\boldsymbol{x})\boldsymbol{M}^{-1}(\xi;\,\boldsymbol{\theta})\boldsymbol{f}(\boldsymbol{x})\right\}\,.
\label{baysdcrit2}
\end{eqnarray}

\noindent For locally D-optimal designs the number of support points $t$ may be between $p$ and $p(p+1)/2+1$. If $t=p$, the optimal design weights are $w_i =1/p$.

\subsection{Design efficiency}
\label{efficsec}

Efficiencies of designs can be compared through the values of the objective function~\eqref{bayscrit}. If $\xi^*_0$ is the optimal design for the prior $p_0(\boldsymbol{\theta})$, comparison is of the values of $\Psi\{\xi^*_0\}$ and of $\Psi\{\xi\}$, where $\xi$ is some other design to be compared and both expectations are taken over the prior $p_0(\boldsymbol{\theta})$.

There is a particularly satisfying form of efficiency for D-optimality when $p(\boldsymbol{\theta})$ is a point prior. Then from~\eqref{baysdcrit} the locally D-optimal design maximizes $|\boldsymbol{M}(\xi;\,\boldsymbol{\theta}_0)|$ and the efficiency of the design $\xi$ is

\begin{equation}
\mbox{Eff}_{\mbox{\scriptsize{D}}} = \left\{\frac{|\boldsymbol{M}(\xi;\,\boldsymbol{\theta}_0)|}{|\boldsymbol{M}(\xi^*_0;\,\boldsymbol{\theta}_0)|}\right\}^{1/p}\,.
\label{deffic}
\end{equation}

\noindent Raising the ratio of determinants to the power $1/p$ results in an efficiency measure which has the dimension of variance; a design with an efficiency of 50\% requires twice as many trials as the D-optimal design to give the same precision of estimation. If only one parameter is of  interest, \eqref{deffic} reduces to comparison of the variances of the estimated parameters under different designs (see \S\S 9.1 and 9.2 of Chapter 1 for a  discussion of efficiency for wider classes of designs and for  functions of parameter estimates including contrasts).

If the prior is not concentrated on a single point, the optimal design has to be found by taking the expectation of the determinant over the prior distribution. Usually this requires a numerical approximation to the value. The efficiency in \eqref{deffic} in addition requires calculation of the expectation of the determinant for the design $\xi$. An informative alternative, which we illustrate in \S\ref{Bayeslogsec}, is to look instead at the distribution of  efficiencies found by simulation from the prior distribution. This procedure avoids taking expectations, since we calculate  \eqref{deffic} for each sampled value of $\boldsymbol{\theta}_0$.

\subsection{Parameter uncertainty}
\label{uncertainsec}

In \S\ref{localbin1vsec} we illustrate the dependence of the locally optimal design on the value of $\boldsymbol{\theta}$ for logistic regression with a single explanatory variable. An example for two-variable logistic regression is in \S\ref{22respsufsec} and for two variable gamma regression in \S\ref{respowerlinksec}. In addition to the Bayesian design of \S\ref{bayesuncsubsec}, which we exemplify in \S\ref{Bayeslogsec}, we here list some other approaches to parameter uncertainty.

\subsubsection{Bayesian designs}
\label{bayesuncsubsec}

Bayesian designs are found to maximize expectations such as~\eqref{bayscrit}. The ease of calculation depends on the form of the prior $p(\boldsymbol{\theta})$ and of $\Psi(.)$ as well as, often crucially,  on the region of integration $\Theta$. The easiest case is when the $\boldsymbol{\theta}$ have a multivariate normal distribution over $\Re^p$, although numerical approximation is still needed. Sometimes a transformation of the parameters is required to achieve this simple structure for multivariate priors. For less amenable cases, a standard solution is to sample from the prior distribution and to use an average objective function. An example for a nonlinear model is in \S18.5 of \citet{ADT:2007}.

Designs maximizing expectations such as~\eqref{bayscrit} ignore the additional effect of the prior information about $\boldsymbol{\theta}$ on the information matrix and make no assumption of a Bayesian analysis. The designs are accordingly sometimes called pseudo-Bayesian. A discussion of Bayesian experimental design is given by \citet{chal+v:95}.

\subsubsection{Sequential designs}
\label{seqsubsec}

Where possible, sequential experimentation can provide an efficient strategy in the absence of knowledge of plausible parameter values. The usual  steps are:

\begin{enumerate}

\item Start with some preliminary information providing an estimate, or guess, of the parameter values. This may lead to either a point prior $\boldsymbol{\theta}_0$ or a prior distribution $p(\boldsymbol{\theta})$.

\item One or a few trials of the optimal design are executed and analysed. If the new estimate of $\boldsymbol{\theta}$ is sufficiently accurate,
the process stops.  Otherwise, step 2 is repeated for the new estimate and the process continued until sufficient accuracy is obtained or the
experimental resources are exhausted.

\end{enumerate}

An early example, for nonlinear regression, is \citet{box+h:65}, extended by \citet[\S17.7]{ADT:2007}. \citet{ds2008} developed a Bayesian sequential design methodology for generalized linear models.

\subsubsection{Minimax and maximin designs}
\label{maximinsubsec}

The minimax approach overcomes dependence of  designs on the unknown value of $\boldsymbol{\theta}$ by finding the best design for the worst case when the parameter $\boldsymbol{\theta}$ belongs to a set $\Theta$. A design $\xi^*$ is found for which

\begin{equation}
\Psi\{\boldsymbol{M}(\xi^*)\} = \underset{\xi}{\min}\;\underset{\boldsymbol{\theta} \in \Theta}
{\max}\;\Psi\{\boldsymbol{M}(\xi;\,\boldsymbol{\theta})\}\,.
\label{maximin}
\end{equation}

In \eqref{maximin} the criterion $\Psi(\cdot)$ needs to be chosen with care. Suppose interest is in D-optimality, with $\xi^*(\boldsymbol{\theta'})$ the locally D-optimal design for parameter value $\boldsymbol{\theta'}$. Unlike with linear models, the value of  $|\boldsymbol{M}\{\xi^*(\boldsymbol{\theta'});\,\boldsymbol{\theta'}\}|$  depends on the value of $\boldsymbol{\theta'}$. Accordingly, maximization of the related design efficiencies  is often used as a criterion, when
the maximin design $\xi^*$ is found to maximize the minimum efficiency:
 $$\mbox{Eff}_{\mbox{\scriptsize{D}}}(\xi^*) =  \underset{\xi}{\max}\;\underset{\boldsymbol{\theta'} \in \Theta}
{\min}\;\mbox{Eff}_{\mbox{\scriptsize{D}}}(\xi;\boldsymbol{\theta'}).$$
\noindent   A potential objection to these designs is that the minimax or maximin design is often close to a combination of locally optimal designs for values of $\boldsymbol{\theta}$ at the edges of the parameter space. If a prior distribution is available, such points may have a very low probability; their importance in the design criterion may therefore be  over-emphasized by the minimax criterion. Providing adequate performance in these unlikely worst case scenarios may greatly affect overall design performance.

A computational difficulty is that such designs can be hard to find. Numerical procedures  are described by \citet{nyquist:2013}, \citet[pp.~82 and p.130]{valerii+sergei:2014} and, in greatest detail, by \citet[\S9.3]{pronz+paz:2013}. Minimax designs for generalized linear models have been found by \citet{sitt:92} and \citet{king+w:2000}. A practical difficulty is that the designs may have an appreciable number of support points, some with very low weights. Of course,  approximations to the optimal design with fewer support points can always be evaluated, provided the optimal design has been found.

\subsubsection{Cluster designs}
\label{clustersubsec}


Some of the computational issues associated with finding Bayesian or minimax designs can be alleviated through the use of clustering of design points, or other less formal techniques, to find designs that incorporate the overall structure of the set of locally optimal designs. Cluster designs are found by (i) sampling parameter vectors $\boldsymbol{\theta}$ from a prior distribution; (ii) finding a locally optimal design for each sampled vector; (iii) clustering the support points of these designs; (iv) forming a new, robust design having equally weighted support points that are the cluster centroids. See \citet{dror+dms:2006}. Such methods are particularly effective in reducing computation when the locally optimal designs are available analytically \citep{russpoiss:2009}.

\subsection{Small effects}
\label{coxsec}

If the effects of the factors are slight, the means  of  observations at different $\boldsymbol{x_i}$ will be similar, and so will each corresponding model weight $u(\boldsymbol{x}_i)$. The information matrix will then, apart from a scaling factor,  be close to the unweighted information matrix $\boldsymbol{X}^\prime\boldsymbol{X}$ and the optimal designs for the weighted and unweighted matrices will be close. Since the designs minimizing functions of $\boldsymbol{X}^\prime\boldsymbol{X}$ are the optimal regression designs, these will be optimal, or near optimal, for generalized linear models with small effects \citep{cox:88}. This is in addition to the result of \S\ref{gamsec} that regression designs are optimal for the gamma model with log link.


\section{Locally optimal designs}
\label{localsec}

\subsection{Binomial data: logistic regression in one variable}
\label{localbin1vsec}

The logistic model is widely used for dose-response data when the response is binomial. For example, \citet{bliss:35} gives data from subjecting  groups of around 60 beetles to eight different levels of an insecticide. The response is the number of beetles killed. The data are reported in numerous textbooks, including \citet{agr:2002}, \citet{coll:2002}, \citet{dobs:2008} and \citet{morg:92}.  The original analysis used a probit model, with more recent analyses preferring a logistic link.  \citet[\S6.14]{a+r:2000} use a goodness of link test to argue that a complementary log-log link is preferable;  \citet[\S22.4]{ADT:2007} present locally optimal designs for both models.

With a single explanatory variable and the logistic link

\begin{equation}
\log\{\mu/(1-\mu)\} = \eta = \theta_0 + \theta_1 x\,.
\label{linlog}
\end{equation}

\begin{figure}[!t]
\centering
\rotatebox{90}{\includegraphics[width=2.4in,clip=true]{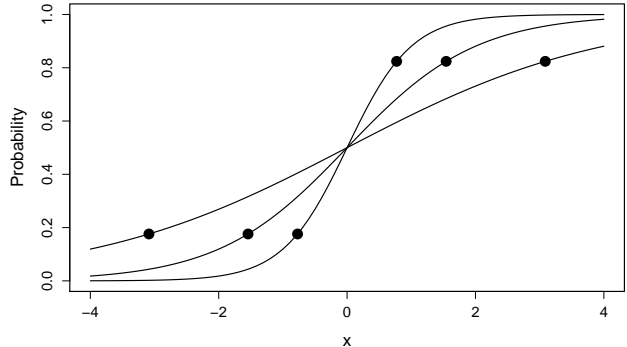}}
\caption{Logistic regression on a single explanatory variable.
Response functions for $\theta_1 = 0.5, 1$ and 2 (the  steepest
curve). The D-optimal design points $\bullet$ are clearly highly
dependent on the value of $\theta_1$. \label{glmf1}}
\end{figure}

\noindent Figure~\ref{glmf1} shows these response curves for $\theta_0 = 0$ and $\theta_1 = 0.5$, 1 and 2. As $\theta_1$ increases, so does the rate of increase of the response at $x = 0$.

It is clear that optimal designs for the three sets of parameter values will be rather different: experiments for values of $x$ for which the responses are virtually all 0 or 1 will be uninformative about the values of the parameters. This intuitive result also follows from the weight $u(x)$~\eqref{logiwt} which goes to zero for extreme values of $\mu$. This result is also related to the phenomenon of separation in the analysis of binary data (see \citealp{firth} and \citealp{woods+:2006}), where a hyperplane of the design space separates the observed data into zeroes and ones.

\begin{figure}[!t]
\centering
\rotatebox{90}{\includegraphics[width=2.4in,clip=true]{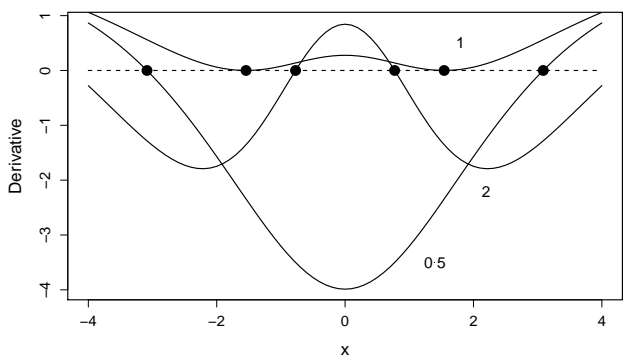}}
\caption{Logistic regression on a single explanatory variable. Equivalence theorem: derivative functions for three D-optimal designs for different values of $\theta_1$ assessed for $\theta_1 = 1$. Curves labelled by value of $\theta_1$; $\bullet$ design points. \label{glmf12}}
\end{figure}

We start by finding the locally D-optimal design for the {\it canonical} case of $\boldsymbol{\theta} = (0,1)^\prime$. From \S\ref{doptsec} we know that the D-optimal design will have either two or three support points. A standard way to proceed is to assume that there are two, when, again from \S\ref{doptsec}, each $w_i = 0.5$. Further, with a response symmetrical about zero, the design can also be assumed symmetrical about zero. It is thus simple to calculate the D-optimal design within this class. The equivalence theorem is then used to check whether the design is indeed optimal. In this case, see Figure~\ref{glmf12}, this is the correct form and
the D-optimal design for a sufficiently large design region $\mathcal X$ is

\begin{equation}
\xi^* = \left\{ \begin{array}{cc} -1.5434 & 1.5434 \\ 0.5 & 0.5
\end{array} \right\}\,,
\label{logoptdes}
\end{equation}

\noindent that is equal weights at two points symmetrical about $x = 0$. At these support points  $\mu$  = 0.176 and 1$-$0.176 = 0.824.

Although designs for other values of the parameters can likewise be found numerically, design problems for a single $x$ can often
be solved in a canonical form, yielding a structure for the designs independent of the particular parameter values (see \S\ref{inducesec}). The translation into a design for other parameter values depends, of course, on the particular $\boldsymbol{\theta}$.

For the upper design point in~(\ref{logoptdes}) the linear predictor $\eta = 0 + 1 \times x $ has the value 1.5434, which is the value we need for the optimal design whatever the parameterization. If we solve~(\ref{linlog}) for the $\eta$ giving this value, the upper support point of the design is given by

\begin{equation}
x^*  = (1.5434 - \theta_0)/\theta_1\,.
\label{xstar}
\end{equation}

For linear regression the D-optimal design puts half of the design weight at the two extreme values of $\mathcal{X}$, whereas, for  logistic regression, the design does not span $\cal X$, provided the region is sufficiently large.  Note that as $\theta_1\rightarrow 0$, the value of $x^*$ increases without limit. This is an example of the result of \citet{cox:88} mentioned above that, for small effects of the variables, the design tends to that for homoscedastic regression. In practice the design region will not be unlimited and, depending on $\boldsymbol{\theta}$, the optimal design may put equal weight on one boundary point and an internal point, or on the two boundary points of $\mathcal{X}$.

\begin{table}
\caption{D-efficiencies (\%), of designs for one-variable logistic regression as $\theta_1$ varies with $\theta_0 = 0$. The design in column $i$ is optimal for the parameter value in row $j$ (100\% efficiency when $i = j$).} \vspace{3mm}
\centering
\begin{tabular}{cccc} & \multicolumn{3}{c}{Design point $x^*$ \eqref{xstar}} \\
$\theta_1$ & 3.086 & 1.5434 & 0.7717  \\ [1ex] \hline \\ [-1.5ex]
0.5        & 100         &  74.52  &  41.52      \\
1          &     57.56   &  100    &  74.52  \\
2          &     5.72    &  57.56  & 100  \\
 [1ex] \hline
\end{tabular}
\label{tabA1}
\end{table}

In addition to the plots of the probability of success $\mu$ against $x$ for three values of $\theta_1$, 0.5, 1 and 2, Figure~\ref{glmf1} also shows the D-optimal design points. For the three parameter values  we obtain design points of $\pm 3.0863$, $\pm1.5434$ and $\pm 0.7717$. The D-efficiencies of each for the other set of parameter values are in Table~\ref{tabA1}. The most striking feature of the table is that the design at $\pm3.0863$, that is for $\theta_1 = 0.5,$ has an efficiency of only 5.72\% when $\theta_1 = 2$. The explanation of this low efficiency is clear from Figure~\ref{glmf1}; the design point is so extreme that the value of $\mu$ when $\theta_1 = 2$ is virtually 0 or 1, so that the experiment yields little information. Clearly we need a compromise design when the parameters are not approximately known. Note that this comparison does not include the value of $\theta_0$, changes in which will make the design asymmetrical about zero. A design robust to the values of $\boldsymbol{\theta}$ typically involves more support points than does the locally optimal design. We give an example in \S\ref{Bayeslogsec}.

We conclude by showing how condition 3 of the Equivalence Theorem in \S\ref{theorsec} can be used both to check the optimality of designs and to suggest improvements to non-optimal designs.

Figure~\ref{glmf12} plots the derivative functions $\psi(x,\xi)$~\eqref{baysdcrit2} for the three designs, assessed for $\theta_1 = 1$. The results therefore relate to the second line of Table~\ref{tabA1}. For $\theta_1 = 1$ the minimum of the function is at zero, the minimum occurring at the points of support of the design. The design is therefore D-optimal.

For the other two designs the minima are both well below zero. The most central set of design points is for $\theta_1 = 2$. Here the minima are around $\pm 2$, indicating that the design points should be more extreme than they are. Likewise, for $\theta_1 = 0.5$, the design points are too extreme and the minimum of the function at the centre of the design region indicates that the design should be shrunk. An important feature of the plot is that, for all three designs, $\psi({x},\xi) = 0$ at the points of support of the design. In order to check the optimality of designs it is necessary to search over $\mathcal{X}$ and determine the minimum of the function, rather than just checking at the design points.

Although we have illustrated the use of $\psi(x,\xi)$ in the assessment of designs, it can also be used in their construction. Sequential addition of design points at the minimum of $\psi(x,\xi)$ leads to the D-optimal design. An example of such sequential design construction is in \S11.2 of \citet{ADT:2007}. A straightforward extension is to the adaptive construction of designs mentioned in \S\ref{seqsubsec} where the parameters are re-estimated as observations become available and one, or a few, trials added at the point maximizing the criterion function. For D-optimality and addition of single observations, this is achieved after $N$ trials by addition of the point minimizing $\psi(x,\xi_N)$ \citep{box+h:65}.

\section{Optimal designs for two or more explanatory variables}
\label{localsevfacsec}

\subsection{Binomial data}
\label{bindes}

\subsubsection{Induced design region}
\label{inducesec}

As the information matrix \eqref{infomat} is of a weighted form, design for the additive linear predictor

\begin{equation}
\eta(\boldsymbol{x}) =  \theta_0  + \sum_{j=1}^{k} \theta_j  x_j\,,
     \label{indes1}
\end{equation}

\noindent is equivalent to (unweighted) design for the linear model

\begin{equation}
E(y) =  \theta_0 \surd u(\boldsymbol{x}) + \sum_{j=1}^{k} \theta_j \surd u(\boldsymbol{x}) x_j,
     = \theta _0 z_0 + \sum_{j=1}^{k} \theta_j z_j\,,
     \label{indes2}
\end{equation}

\noindent where $u(\boldsymbol{x})$  is defined in \eqref{eq6}.
Hence the original design region $\mathcal{X}$ can be transformed to the induced design region $\mathcal{Z}$ for the induced variables $z_0,\ldots,z_{k}$. Clearly, $\mathcal{Z}$ depends on both $\mathcal{X}$ and $\boldsymbol{\theta}$.

\citet{ftw:92} used this relationship with linear model design to provide geometric insight into the structure of designs for single variable generalized linear models. With one explanatory variable, $\mathcal{X}$ is a line segment $ a \le x \le b$. However, because in~\eqref{indes2} $z_0 = \surd u(x)$, $\mathcal{Z}$ is of dimension 2. For the models for binomial data of \S\ref{binsec} and $x\in\mathbb{R}$, $\mathcal{Z}$ is a closed curve similar to the ``design locus" in Figure~2 of \citet{box+l:59}, the exact shape of which will vary with $\boldsymbol{\theta}$.  The results of
\citet{ftw:92} require the careful enumeration of a large number of criteria and special cases. For  D-optimality they use results on the relationship between D-optimal designs and the ellipse of minimum volume centered at the origin that contains $\mathcal{Z}$ \citep{s+dmt:73,sib:74,silv:80}.

\citet{ftw:92} are concerned with exponential family linear models with a single explanatory variable. \citet{Wu01062014} provide results for a single-variable  model with a quadratic predictor.  Mathematical results for the linear predictor~\eqref{indes2} with more than one explanatory variable are not generally available. An important limitation on the class of models for which results can be expected comes from the form of~\eqref{indes2} which excludes interaction and higher-order polynomial terms. We have written $z_j = \surd u(\boldsymbol{x}) x_j$ and so $z_k = \surd u(\boldsymbol{x}) x_k$. But $z_jz_k  \ne u(\boldsymbol{x}) x_jx_k$.

In \S\ref{binsevfac} we compute locally optimal designs for binomial data with linear predictor~\eqref{indes1} for two variables. Views  of the designs in $\mathcal{X}$ and in  $\mathcal{Z}$ are quite distinct, but both are informative about the structure of the designs. Those in $\mathcal{X}$ relate the design points to the values of the response, whereas those in $\mathcal{Z}$ show that the design points are at the extremes of the region and, for second-order models, near centres of edges and at the centre of the region. The relationship with response surface designs is clear, even if the mirror is distorted.

\subsubsection{First-order models}
\label{binsevfac}

The properties of designs for response surface models, that is with two or more continuous explanatory variables, depend much more on the
experimental region than those where there is only one factor.

Although it was assumed in the previous section that the experimental region $\mathcal{X}$  was effectively unbounded, the design was constrained by the weight $u$ to lie in a region in which $\mu$ was not too close to zero or one. But with more than one explanatory variable, constraints on the region are necessary. For example, for the two-variable first-order model

\begin{equation}
\log\{\mu/(1-\mu)\} = \eta = \theta_0 + \theta_1 x_1 + \theta_2 x_2\,,
\label{22twofac}
\end{equation}

\noindent with $\boldsymbol{\theta}^\prime = (0, \gamma, \gamma)$, all design points for which $x_1+x_2= 0$ yield a value of 0.5 for $\mu$, however extreme the values of
$x$.

\begin{table}[!t]
\begin{center}
\caption{Sets of parameter values for first-order linear predictors in two variables with the logistic link.} \label{22tabB1} \vspace{5mm}
\begin{tabular}{c|ccc}
Set & $\theta_0$ & $\theta_1$  & $\theta_2$ \\
\hline
       &           &            &           \\
    B1 &       0   &          1 &         1 \\
    B2 &       0   &          2 &         2 \\
    B3 &       2   &          2 &         2 \\
    B4 &       2.5 &          2 &         2
  \end{tabular}
  \end{center}
  \end{table}

We now explore designs for the linear predictor~\eqref{22twofac} with the logistic link for a variety of parameter values and $\mathcal{X}=[-1,1]^2$. These and succeeding designs were found by a numerical search with a quasi-Newton algorithm. The constraints to ensure that $\boldsymbol{x} \in {\cal X}$ and on the design weights were enforced using the trigonometric transformations listed in \citet[\S9.5]{ADT:2007}.

Four sets of parameter values are given in Table~\ref{22tabB1}. D-optimal designs for the sets B1 and B2 are listed in Table~\ref{22tabB2}. The parameter values of B1 $(0,1,1)$ are closest to zero. The table shows that the design has support at the points of the $2^2$ factorial, although the design weights are not quite equal. They are so for the normal theory model and become so for the logistic model as $\theta_1$ and $\theta_2$ $\rightarrow 0$ with $\theta_0 > 0$. At those factorial points for which $x_1 + x_2 = 0, \mu = 0.5$ since $\theta_1 = \theta_2$. At the other design points $\mu$ = 0.119 and 0.881, slightly more extreme values than the values of 0.176 and 0.824 for the experiment with a single variable.

\begin{table}[!t]
\begin{center}
\caption{D-optimal designs for logistic models with the sets of parameter values B1 and B2 of Table~\ref{22tabB1}; $w_i$ design weights}
\label{22tabB2} \vspace{5mm}
\begin{tabular}{r|rrrr||rrrrr}
     & \multicolumn{4}{c||}{Design for B1}
     & \multicolumn{5}{c}{Design for B2 }  \\
 $i$ & \multicolumn{1}{c}{$x_{1i}$}
     & \multicolumn{1}{c}{$x_{2i}$}
     & \multicolumn{1}{c}{$w_i$ }
     & \multicolumn{1}{c||}{$\mu_i$}
     & \multicolumn{1}{c}{$x_{1i}$}
     & \multicolumn{1}{c}{$x_{2i}$}
     & \multicolumn{1}{c}{$w^{(1)}_i$}
     & \multicolumn{1}{c}{$w^{(2)}_i$}
     & \multicolumn{1}{c}{$\mu_i$} \\
\hline
   &    &    &       &       &         &         &       &       &       \\
 1 & $-$1 & $-$1 & 0.204 & 0.119 &  0.1178 & $-$1.0000 &       & 0.240 & 0.146 \\
 2 &    &    &       &       &  1.0000 & $-$0.1178 &       & 0.240 & 0.854 \\
 3 &  1 & $-$1 & 0.296 & 0.500 &  1.0000 & $-$1.0000 & 0.327 & 0.193 & 0.500 \\
 4 & $-$1 &  1 & 0.296 & 0.500 & $-$1.0000 &  1.0000 & 0.193 & 0.327 & 0.500 \\
 5 &    &    &       &       & $-$1.0000 &  0.1178 & 0.240 &       & 0.146 \\
 6 &  1 &  1 & 0.204 & 0.881 & $-$0.1178 &  1.0000 & 0.240 &       & 0.854
\end{tabular}

  \end{center}
  \end{table}



An interesting feature of our example is that the number of support points of the design depends upon the values of the parameter $\boldsymbol{\theta}$. From the discussion of Carath\'eodory's Theorem  in \S\ref{theorsec}, the maximum number of support points required by an optimal design is usually $p(p+1)/2$ \citep[\S8.3]{puke:93}. Our
second set of parameters, B2 in which $\boldsymbol{\theta}^\prime = (0,2,2)$, gives two four-point optimal designs, with weights given by $w^{(1)}_i$ and $w^{(2)}_i$ in Table~\ref{22tabB2} and support points where $\mu = 0.146$, $0.5$ and $0.854$.  Any convex combination of these two designs, $\alpha w^{(1)}_i + (1-\alpha)w^{(2)}_i$ with $0\leq\alpha\leq 1$, will also be optimal, and will have six support points, which is the value of the usual bound when $p =3$. These two component designs arise from the symmetry of the design problem; not only does $\theta_1 = \theta_2$ but also the design region is  symmetrical in $x_1$ and $x_2$.

\begin{table}
\begin{center}
 \caption{D-optimal designs for logistic models with the parameter values
 B3 and B4 of Table~\ref{22tabB1}; $w_i$ design weights}
\label{22tabB3} \vspace{5mm}
\begin{tabular}{r|rrrr||rrrr}
     & \multicolumn{4}{c||}{Parameter set B3}
     & \multicolumn{4}{c}{Parameter set B4}   \\
 $i$ & \multicolumn{1}{c}{$x_{1i}$}
     & \multicolumn{1}{c}{$x_{2i}$}
     & \multicolumn{1}{c}{$w_i$ }
     & \multicolumn{1}{c||}{$\mu_i$}
     & \multicolumn{1}{c}{$x_{1i}$}
     & \multicolumn{1}{c}{$x_{2i}$}
     & \multicolumn{1}{c}{$w_i$}
     & \multicolumn{1}{c}{$\mu_i$} \\
\hline
  &          &         &       &       &         &         &       &       \\
1 &  $-$1.0000 & $-$0.7370 & 0.169 & 0.186 & $-$1.0000 &  0.5309 & 0.333 & 0.827 \\
2 &  $-$1.0000 &  0.7370 & 0.331 & 0.814 & $-$1.0000 & $-$1.0000 & 0.333 & 0.182 \\
3 &  $-$0.7370 & $-$1.0000 & 0.169 & 0.186 &  0.5309 & $-$1.0000 & 0.333 & 0.827 \\
4 &   0.7370 & $-$1.0000 & 0.331 & 0.814 &         &         &       &
\end{tabular}
  \end{center}
  \end{table}

\begin{figure}[!b]
   \begin{center}
   \includegraphics[width=5.0in,clip=true]{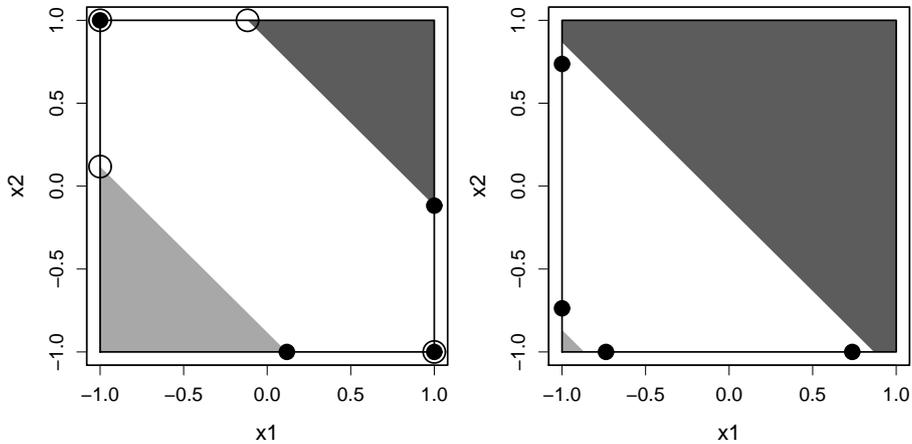}
   \caption{Support points for D-optimal designs for logistic models with parameter values
   B2 and B3
   in Table~\ref{22tabB1}.  In the lightly shaded area $\mu \leq
   0.15$, whereas, in the darker region, $\mu \geq 0.85$.  In the left-hand panel, one
    four-point design for B2 is depicted by open circles, the other by filled circles. The symmetry of the designs is evident.}
   \label{figb34}
   \end{center}
\end{figure}

The D-optimal designs for the two remaining sets of parameters in Table~\ref{22tabB1} are given in Table~\ref{22tabB3}. These
designs have respectively 4 and 3 points of support. When $\boldsymbol{\theta}^\prime = (2,2,2)$, the design points are where $\mu = 0.186$ and 0.814. For $\boldsymbol{\theta}^\prime = (2.5,2,2)$ the minimum value of $\mu$ is 0.182 at $(-1, -1)$ and the experimental values of $\mu$ are  0.182 and 0.827. For this three-point design for a three parameter model, the design weights $w_i = 1/3$.  A useful general indication is that an informative experiment should have $0.15 < \mu < 0.85$. This bound is included in the plots of designs in Figure~\ref{figb34}.



The relationship between the design points  and the values of $\mu$ are shown, for parameter values B2 and B3, in Figure~\ref{figb34}. For $\boldsymbol{\theta}^\prime = (0,2,2)$, one     four-point design for B2 is depicted by open circles, the other by filled circles; the symmetry of the designs is evident. For $\boldsymbol{\theta}^\prime = (2,2,2)$ there are again four support points of the design, which now lie somewhat away from the boundaries of the regions of high and low values of $\mu$.

\subsubsection{Induced design region for the logistic link}
\label{22inducesec}

\begin{figure}
   \begin{center}
   \includegraphics[width=5.0in,clip=true]{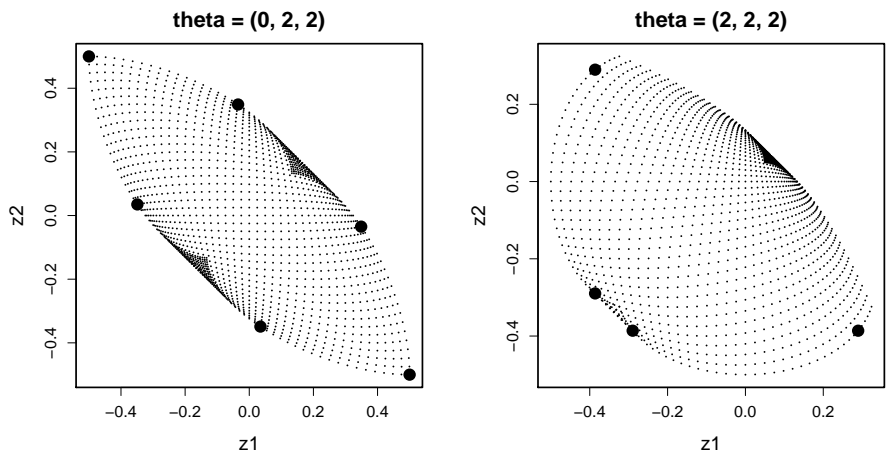}
   \caption{Support points for D-optimal designs for logistic models with parameter values
   B2 (left) and B3 (right) of Table~\ref{22tabB1} in the induced design region $\mathcal{Z}$. For these first-order models all design points lie on the boundary of $\mathcal{Z}$.}
   \label{fig22b3}
   \end{center}
\end{figure}

For the first-order model in $k=2$ factors \eqref{22twofac}, for which $p=3$, the induced design space $\mathcal Z$ is of dimension three. Two examples, projected onto $z_1$ and $z_2$ and thus ignoring $z_0 = \surd u(\boldsymbol{x})$, are given in Figure~\ref{fig22b3} for $\mathcal{X}$ the unit square. In the left-hand panel of the figure, $\boldsymbol{\theta}^\prime = (0,2,2)$ so that at the corner of $\mathcal X$ for which $x_1 = x_2 = 1$, $\eta = 4$ and $\mu = 0.982$. This is well beyond the range for informative experiments and the projection of the induced design space appears to be folded over. As a consequence, experiments at extreme positions in $\mathcal Z$ are not at extreme points in $\mathcal X$. The results in the other panel for $\boldsymbol{\theta}^\prime = (2,2,2)$ are similar, but more extreme. For both sets of parameter values the design points lie, as they should, on the boundary of the induced design region.

These examples show the importance of both the design region and the value of $\mu$ in determining the optimal design. In order to reveal the structure of the designs as clearly as possible, the designs considered have all had $\theta_1 = \theta_2$, and so are symmetrical in $x_1$ and $x_2$.  When $\theta_1\ne\theta_2$, both the design region and the values of $\mu$ are important in the resulting asymmetrical designs. Asymmetrical designs also arise when the log-log and complementary log-log links are used, since these links are not symmetrical functions.

\subsubsection{Theoretical results for a first-order predictor}
\label{binfotheory}

Various authors have derived optimal designs for $k>1$ for first-order linear predictor~\eqref{indes1} for some special cases; see \citet{sitt+bent:95}, \citet{sitt:95} and \citet{tors+g:2001}. \citet*{yzh2011} considered the case when the assumption of a bounded design region $\mathcal{X}$ is relaxed by allowing one variable, say $x_k$, to take any value in $\Re$. As the D-optimality objective function is still bounded in this case, the authors were able to provide a methodology to obtain a locally D-optimal design analytically. We restate their corollary~1 for the special case of logistic regression and D-optimality.

\vspace{0.5cm}

\textbf{Theorem~1}: \citep{yzh2011} \textit{For the logistic model with linear predictor~\eqref{indes1} and $\mathcal{X}=[-1,1]^{k-1}\times\Re$, a D-optimal design has $2^k$ equally weighted support points,}

$$
\boldsymbol{x}_{l}^\star = \left\{
\begin{array}{ll}
(x_{1l},\ldots,x_{(k-1)l},a_{l}^\star) & \mbox{for } l=1,\ldots,2^{k-1} \\
(x_{1l},\ldots,x_{(k-1)l},-a_{l}^\star) & \mbox{for } l=2^{k-1}+1,\ldots,2^{k}\,, \\
\end{array}
\right.
$$

\noindent \textit{where}

$$
x_{jl} = \left\{
\begin{array}{rl}
-1 & \mbox{if } \lceil\frac{l}{2^{k-1-j}}\rceil \mbox{ is odd}\\
1 & \mbox{if } \lceil\frac{l}{2^{k-1-j}}\rceil \mbox{ is even}
\end{array}
\right.
j=1,\ldots,k-1\,,
$$

\noindent \textit{$\lceil a\rceil$ is the smallest integer greater than or equal to $a$, the numerator of the fractions is l and $a_{l}^\star = \eta^\star - \eta_{l}^\star(k)$. Here, $\eta^\star$ maximises}

$$
\eta^2\left\{\frac{dh(\eta)/d\eta}{h(\eta)[1-h(\eta)]}\right\}^{k+1}\,,
$$

\noindent \textit{where $h=g^{-1}$, the inverse of the logistic link function, and $\eta_l^\star(k) = \theta_0 + \sum_{j=1}^{k-1}\theta_jx_{jl}$.}

\vspace{0.5cm}

This result has a fairly straightforward interpretation. If we fix the values of $k-1$ variables in the support points at the level combinations of a $2^{k-1}$ factorial design, then the selection of an optimal design reduces to a one variable problem (the choice of the values to be taken by $x_k$). Note that ${\cal X}$ is such that each variable lies in an interval, rather than just taking the two values $\pm 1$.

To illustrate this result, Table~\ref{yangtable} gives D-optimal designs on $[-1,1]\times\Re$ for the sets of parameter values in Table~\ref{22tabB1}. These designs are quite different from the designs in Tables~\ref{22tabB2} and~\ref{22tabB3}, where the restriction of $x_2$ to the interval $[-1,1]$ results in the designs having different numbers of support points, different values for the supports points and different weights.

\begin{table}
\caption{\label{yangtable}D-optimal support points from Theorem~1 \citep{yzh2011} for parameter values in Table~\ref{22tabB1}. For each design, the support points are equally weighted.}
\begin{center}
\begin{tabular}{rrr}
\hline
 & \multicolumn{2}{c}{Support points} \\
Parameters & $x_1$ & $x_2$ \\
\hline
B1 & $-1$ & $-2.2229$ \\
 & $-1$ & $2.2229$ \\
 & $+1$ & $-0.2229$ \\
 & $+1$ & $0.2229$\\
B2 & $-1$ & $1.6115$\\
 & $-1$ & $-1.6115$\\
 & $+1$ & $-0.3886$\\
 & $+1$ & $0.3886$\\
B3 & $-1$ & $0.6115$\\
 & $-1$ & $-0.6115$\\
 & $+1$ & $-1.3886$\\
 & $+1$ & $1.3886$\\
B4 & $-1$ & $-0.3615$\\
 & $-1$ & $0.3615$\\
 & $+1$ & $-1.6386$\\
 & $+1$ & $1.6386$\\
\hline
\end{tabular}
\end{center}
\end{table}

\subsubsection{A second-order response surface}
\label{22respsufsec}

\begin{figure}
   \begin{center}
   \includegraphics[width=4.5in,clip=true]{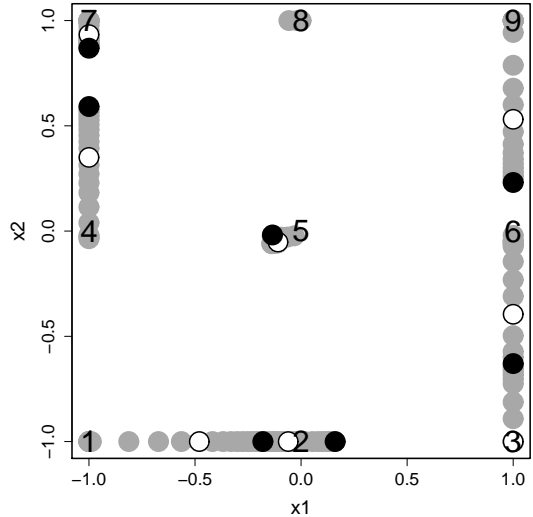} \vspace{-5mm}
   \caption{D-optimal designs for second-order logistic model as $\gamma$ varies ($0 \le \gamma
   \le 2$).  Support points: numbers, $\gamma = 0$; circles, $\gamma =
   1$; black dots $\gamma = 2$; and grey dots, intermediate values.}
   \label{fig22b4}
   \end{center}
   \end{figure}

This section extends the results of earlier sections to the second-order response surface model, again with two factors and again with the
logistic link.  The purpose  is to show the relationship with, and differences from, designs for regression models.
The D-optimal designs are found, as before, by minimizing $-\log |\boldsymbol M(\xi;\,\boldsymbol{\theta})|$. With $n =9$ and two-dimensional $\boldsymbol{x}$, the numerical search is in 26 dimensions. However, once the structure of the designs is established, with 8 of the design points on the edges of ${\cal X}$ (see Figure~\ref{fig22b4}) the search can be reduced to 18 dimensions, with the equivalence theorem providing a check on this reduction.

To explore how the design changes with the parameters of the model we look at a series of designs for the family of linear predictors
\begin{equation}
\eta = \theta_0 + \gamma( \theta_1x_1 +
\theta_2x_2 +  \theta_{12}x_1x_2 + \theta_{11}x_1^2 + \theta_{22}x_2^2)
\;\;\;\mbox{with} \;\;\;\gamma \ge 0\,,
\label{22linpred}
\end{equation}
\noindent and design region $\mathcal{X}=[-1,1]^2$. The parameter $\gamma$ provides a family of similarly shaped linear predictors  which increasingly depart, in a proportional way,  from a constant value  as $\gamma$ increases. When $\gamma = 0$ the result of \cite{cox:88} shows that the design is the D-optimal design for the second-order regression model, the unequally weighted $3^2$ factorial design given in \S\ref{respowerlinksec}.

For numerical exploration we take $\theta_0 = 1$, $\theta_1 =  2$, $\theta_2 =2$, $\theta_{12} =   -1$, $\theta_{11} =  -1.5 $ and $\theta_{22} =  1.5$. As $\gamma$ varies from 0 to 2, the shape of the response surface becomes increasingly complicated.

Figure~\ref{fig22b4} shows the support points of the D-optimal designs as $\gamma$ increases from zero in steps of 0.1. The design points are labelled, for $\gamma = 0$, in standard order for the $3^2$ factorial, with $x_1$ changing more frequently. The figure shows how all but one of the design points stay on the boundary of the design region; the circles and black dots are the support points for $\gamma = 1$ and $2$, respectively, with the grey dots indicating intermediate values. There is little change in the location of the centre point, point 5, over this range of values for $\gamma$. Initially the design has nine points, but the weight on point 8 decreases to zero when $\gamma = 0.3$. Thereafter, the design has eight support points until $\gamma = 1.4$ when the weight on observation 6 becomes zero.

\begin{figure}
   \begin{center}
   \includegraphics[width=5.0in,clip=true]{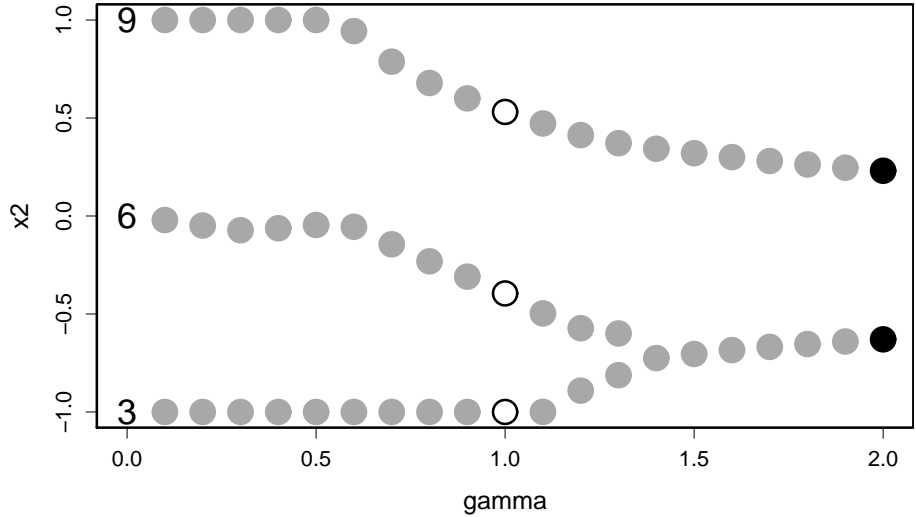}
   \caption{D-optimal designs for second-order logistic model. The values of $x_2$ for support points 3, 6 and 9 of Figure~\ref{fig22b4}  as $\gamma$ varies between zero and two. The coding of the symbols is the same as in the earlier figure.}
   \label{fig22b5}
   \end{center}
\end{figure}


Figure~\ref{fig22b5} serves to help interpret the behaviour of the design as $\gamma$ increases, showing the values of $x_2$ for the three design points (3, 6 and 9) in Figure~\ref{fig22b4} for which $x_1 = 1$. Initially the values of $x_2$ are those for the $3^2$ factorial and they remain virtually so until $\gamma = 0.6$. Thereafter, they gradually converge towards three more central values.

\begin{figure}
   \begin{center}
   \pdfimageresolution=600
   \includegraphics[width=5.2in,clip=true]{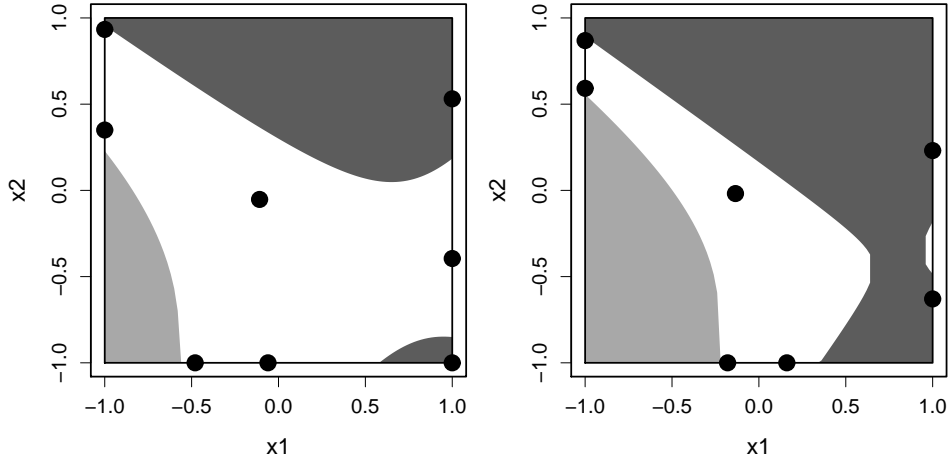}
   \caption{Support points for D-optimal designs for the second-order logistic model. Left-hand panel $\gamma = 1$, right-hand panel $\gamma =
   2$. In the lightly shaded area $\mu \le 0.15$, whereas, in the
   darker region, $\mu \ge 0.85$.}
   \label{fig22b6}
   \end{center}
\end{figure}

The relationship between the support points of the design and the values of $\mu$ is highlighted in Figure~\ref{fig22b6} where, as in Figure~\ref{figb34}, the pale areas are regions in which $\mu \le 0.15$, with the dark regions the complementary ones where $\mu \ge 0.85$. The left-hand panel of Figure~\ref{fig22b6}, for $\gamma = 1$, shows that the 8-point design is a distortion of a standard response surface design, with most points in the white area and the remainder on the boundary of the design region, some close to the contours of $\mu$ = 0.15 or 0.85. In the numbering of
Figure~\ref{fig22b4}, points 2 and 6 are on the edge of the design region where $\mu$ is close to 0.5. Points 3 and 9 are at higher values of $\mu$.

A similar pattern is clear in the seven-point design for $\gamma =2$ in the right-hand panel of the figure; four of the seven points are on the edges of the white region, one is in the centre and only points 3 and 9 are at more extreme values of $\mu$.

The two panels of Figure~\ref{fig22b6} taken together explain the trajectories of the points in Figure~\ref{fig22b4} as $\gamma$ varies. For example, points 1 and 4 move away from $(-1,-1)$ as the value of $\mu$ at that point decreases, point 3 remains at $(1, -1)$ until $\gamma$ is
close to one and point 8 at $(0, 1)$ is rapidly eliminated from the design as the value of $\mu$ there increases with $\gamma$.

Further insight into the structure of the designs can be obtained from consideration of the induced design region introduced in \S\ref{22inducesec}. Although, as stated earlier, extension of the procedure based on~\eqref{indes2} to second-order models such as~\eqref{22linpred} is not obvious, it is still informative to look at the plot the designs in $\mathcal{Z}$ space. The left-hand panel of Figure~22.8 of \citet{ADT:2007} shows the eight-point design for $\gamma = 1$ plotted against $z_1$ and $z_2$; seven points lie on the edge of this region, well spaced and far from the centre, which is where the eighth point is. The right-hand panel for $\gamma = 2$ shows six points similarly on the edge of $\mathcal{Z}$; the centre point is hidden under the seemingly folded-over region near the origin.

In the induced design region these designs are reminiscent of response surface designs, with a support point at the centre of the
region and others at remote points. However the form of $\mathcal Z$ depends on the unknown parameters of the linear predictor, so this
description is not helpful in constructing designs. In the original space $\mathcal X$ we have described the designs for this second-order model  as a series of progressive distortions of designs with support at the points of the $3^2$ factorial.  For small values of $\gamma$ the unweighted $3^2$ factorial provides an efficient design, with a D-efficiency of 97.4\% when $\gamma = 0$. However, the efficiency of this design declines steadily with
$\gamma$, being 74.2\% for $\gamma = 1$ and a low 38.0\% when $\gamma = 2$. If appreciable effects of the factors are expected,
the special experimental design methods of this section need to be used. Further discussion of designs for two variable logistic models is given by \citet{sitt+bent:95} with particular emphasis on the structure of ${\cal Z}$.

\subsubsection{Bayesian D-optimal designs}
\label{Bayeslogsec}

We can also find Bayesian D-optimal designs, maximizing~\eqref{baysdcrit}, for response surface designs and binomial data. Motivated by a food technology example, \citet{woods+:2006} found designs for logistic regression with linear predictor

$$
\eta(\boldsymbol{x}) = \theta_0 + \sum_{i=1}^3\theta_ix_i +\sum_{i=1}^3\sum_{j\ge i}^3 \theta_{ij}x_ix_j \,,
$$

\noindent with $x_i\in[-1.2782,1.2782]$. Here we find a Bayesian D-optimal design assuming independent uniform prior densities for the parameters in $\boldsymbol{\theta}$, defined on the support

$$
\theta_1,\theta_2\in [2,6]\,,\quad \theta_0,\theta_3,\theta_{ij}\in [-2,2]\, \mbox{ for } i,j = 1,2,3\,.
$$

\noindent We approximate the expectation~\eqref{baysdcrit} by the sample average across a 20 run Latin Hypercube sample (see, for example, Chapter~19 and \citealp{bnoz+:2003}, Ch. 6). \citet{brad+got:2009} and \citet{wv} discuss and compare some alternative approaches for this approximation.

\begin{figure}[t]
\begin{center}
\includegraphics[scale=0.5]{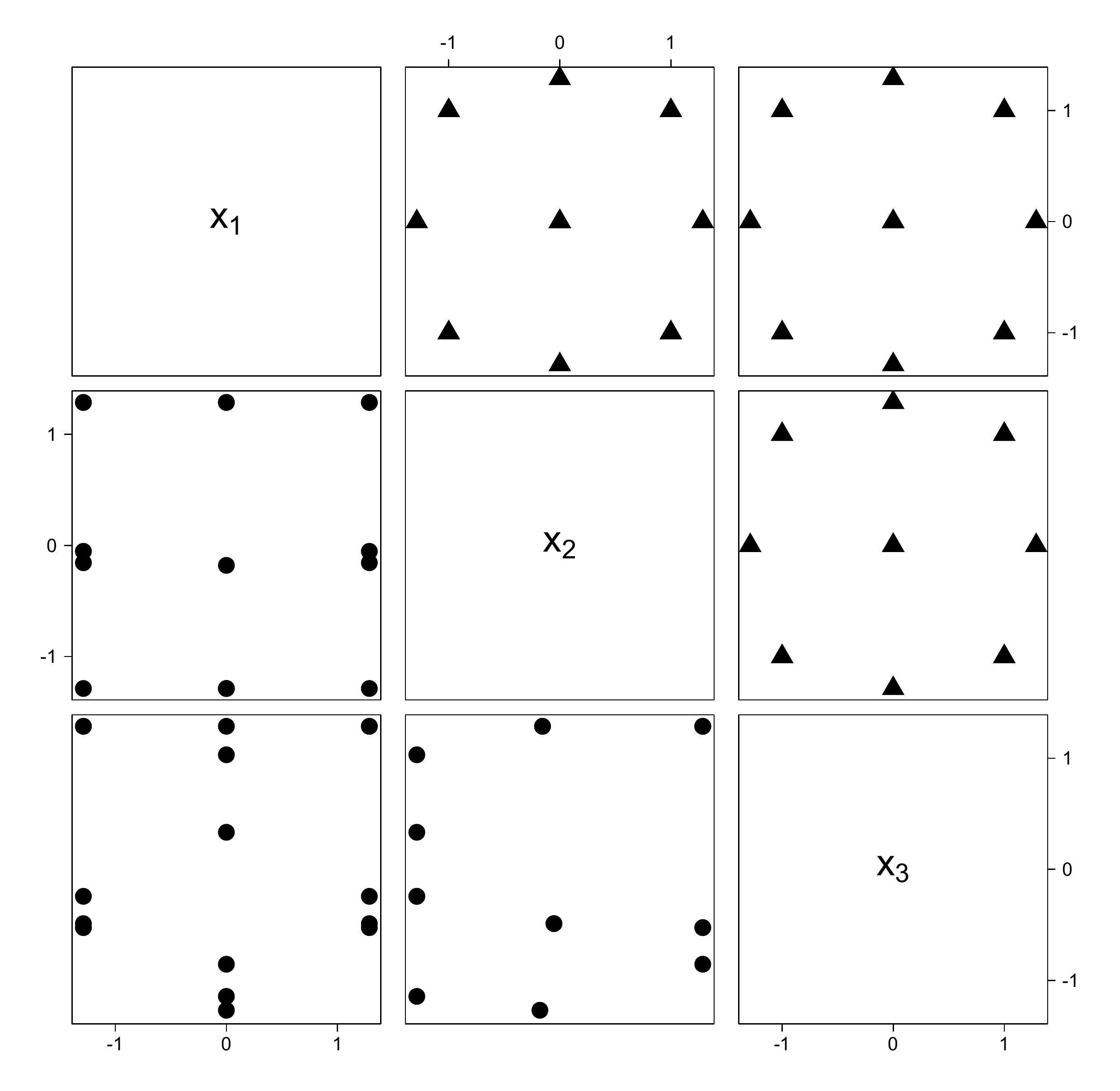}
\end{center}
\caption{\label{binbayesdes}Two-dimensional projections of the Bayesian D-optimal design for a logistic regression model (lower diagonal) and a central composite design (upper diagonal).}
\end{figure}

For this example, a simulated annealing algorithm \citep{woods2010} was used to find an exact design, $d_{16}$, with $n=16$ points. The design is given in Figure~\ref{binbayesdes} and, in fact, has $t=n=16$ support points and no replication. For reference, Figure~\ref{binbayesdes} also gives the design points of a 16 run central composite design (CCD, see Chapter~10), $d_{ccd}$, with 8 factorial points, 6 axial points with $x_j = \pm 1.2782$ and two centre points; see, for example, \citet{box+d:2007}. This design is standard for normal theory response surface studies and is an obvious comparator for the Bayesian GLM design. In fact, a CCD had been employed for the food technology example in previous experimentation.

While there are some familiar features to the Bayesian D-optimal design, including (near) centre points, there are also some distinct differences from the CCD. These include the presence of extreme corner points in the D-optimal design and, for $x_1$ and $x_2$, fewer distinct levels ($\sim 3$ for each of these two variables).

\begin{figure}[t]
\begin{center}
\includegraphics[scale=0.65]{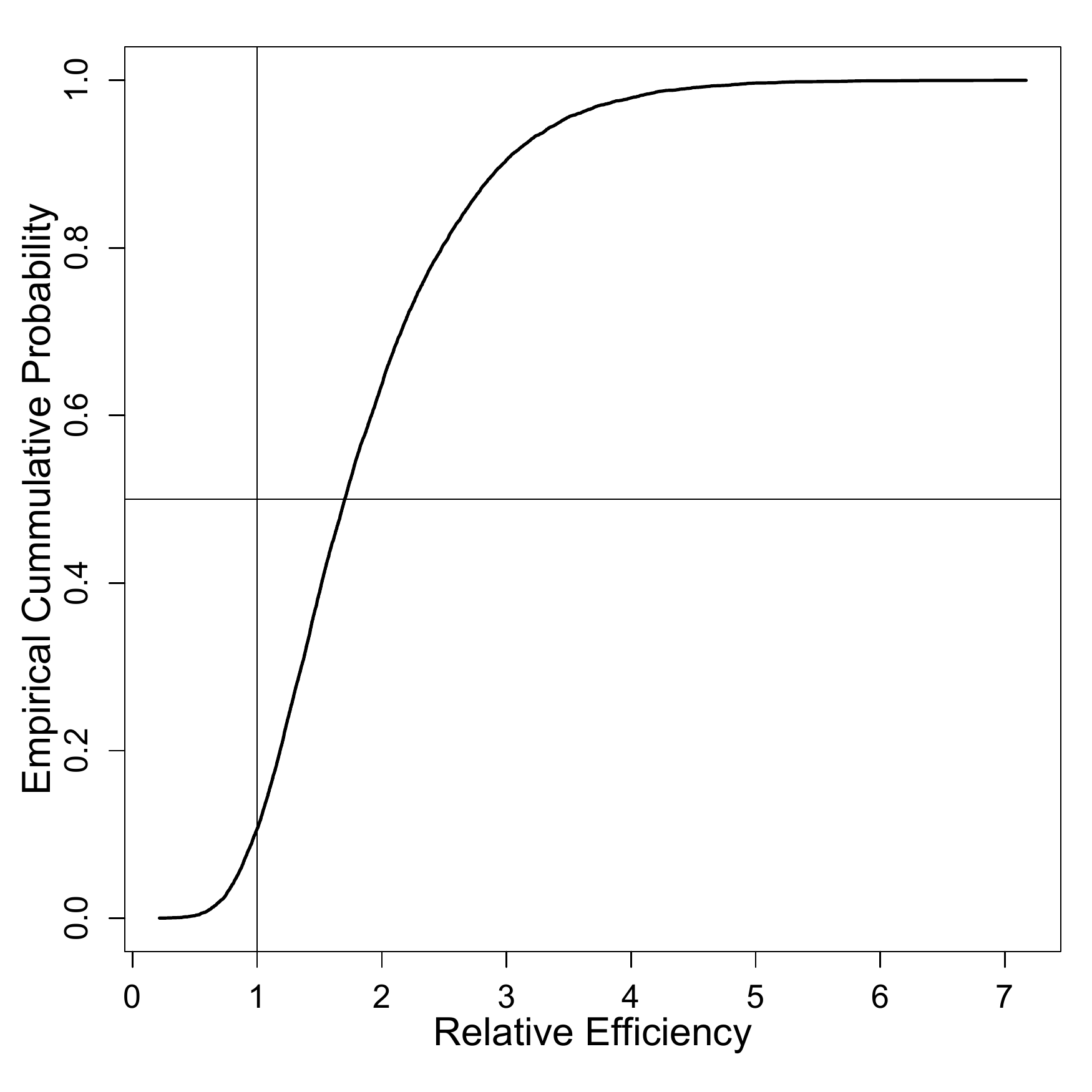}
\end{center}
\caption{\label{binbayeseff} Empirical cumulative distribution function of the relative D-efficiency of the Bayesian D-optimal design compared to the central composite design.}
\end{figure}

The relative performance of the designs was assessed via a simulation study. A sample of 1000 parameter vectors, $\boldsymbol{\theta}^{(l)}$, was drawn from the prior distribution for $\boldsymbol{\theta}$. For each vector, we calculated the relative D-efficiency

\begin{equation}\label{releff}
\mbox{Eff}_D(l) = \left\{
\frac{|\boldsymbol{M}(d_{16});\,\boldsymbol{\theta}^{(l)}|}
{|\boldsymbol{M}(d_{ccd});\,\boldsymbol{\theta}^{(l)}|}
\right\}^{\frac{1}{10}}\,,\qquad l=1,\ldots,1000\,.
\end{equation}

The empirical cumulative distribution of the relative efficiency, Figure~\ref{binbayeseff}, shows a dramatic difference in performance between the two designs. The Bayesian D-optimal design is more efficient than the CCD for about 85\% of the sampled parameter vectors, and has efficiency of 1.75 or more for 50\% of the sampled vectors. The maximum relative efficiency is close to 7.

\subsection{Gamma data}
\label{gamdatsec}

\subsubsection{Theoretical results for first-order designs with the power link}

\label{powerlinksec}

We use numerical examples to illustrate the great difference between designs for gamma models and those for binomial responses of \S\ref{binsevfac}. The examples are calculated using the power link. In this introductory section we present the theoretical results of \citet{burr+s:94} for first-order models. Our numerical example of a first-order model is in \S\ref{numpowerlinksec} with an example for a response-surface model in \S\ref{respowerlinksec}. We conclude with a brief comparison of designs for the power link with those for the Box-Cox link.

 If the design region for uncoded variables is of the form $a_j \le x_j^u \le b_j$, the   region can be coded so that $0 \le x_j \le 1$ for $j = 1,\ldots,k$. The requirement that $\mu > 0$ for all non-negative $\boldmath{x}$ leads to a canonical form of the original problem  with $\theta_j \ge 0$ for all $j = 0,\ldots,k$, with at least one inequality.  Since the weights \eqref{powerlinku} are monotonic in $\eta$, the support points of D-optimal designs as $\boldsymbol{\theta}$ varies must then be some of the points of the $2^k$ factorial. Which points have non-zero weights depends on the values of the $\theta_j$. For effects large relative to $\theta_0$, \citet{burr+s:94} provide the following theorem.

\vspace{0.5cm}

{\bf Theorem~2}. \citep{burr+s:94} \textit{For the coded design variables x$_j$, the design which puts weights 1/(k+1) at each of the k+1 points }
\[ (0,\ldots,0)', (1,0,\ldots,0)',(0,1,0,\ldots,0)',\ldots,(0,\ldots,0,1 )'  \]
\textit{is D-optimal for gamma regression, the power link and a first-order linear predictor if, and only if, for all $i,j = 1,\ldots,k$,}
\[\theta_0^2 \le \theta_i\theta_j. \]

Thus, for `large' effects a `one-factor-at-a-time' approach is optimal. However, as the effects become smaller, the design approaches the $2^k$ factorial in line with the result of \citet{cox:88} discussed in \S\ref{coxsec}. Of course, the weights $w_i$ have to be found numerically. However, the numerical search is greatly simplified by being restricted to the support of the $2^k$ factorial. It is also a great simplification that, as we showed in  \S\ref{gamsec}, the designs do not depend on the value of the power $\kappa$.

\subsubsection{Examples of first-order designs with the power link}
\label{numpowerlinksec}

In both examples there are two explanatory variables. We only look at symmetrical designs generated with  $\theta_1 = \theta_2$ having the three values 0.1, 0.5 and 1. In all calculations $\theta_0 = 1$. The resulting designs are in Table~\ref{gamtab1}.

\begin{table}[!t]
\caption{D-optimal designs for two-variable first-order model for gamma responses with the power link; $\boldsymbol{\theta} = (1,\chi,\chi)'$. } \vspace{3mm}
\centering
\begin{tabular}{ccccc} &  & \multicolumn{3}{c}{Design weights $w_i$} \\
$x_1$ & $x_2$ & $\chi = 0.1$ & $\chi = 0.5$ & $\chi = 1$  \\ [1ex] \hline \\ [-1.5ex]
0     & 0  &  0.271  &  0.313  & 1/3 \\
0     & 1  &  0.252  &  0.281  & 1/3 \\
1     & 0  &  0.252  &  0.281  &  1/3 \\
1     & 1  &  0.225  & 0.125   & 0 \\
 [1ex] \hline
\end{tabular}
\label{gamtab1}
\end{table}

These results nicely illustrate the theoretical results of \S\ref{powerlinksec}. We have parameterized the problem with $\theta_1 = \theta_2 = \chi$. For $\chi = 0.1$, that is with small effects, the design is virtually the $2^2$ factorial, with weights close to 1/4 ranging from 0.225 to 0.271. Increasing $\chi$ to 0.5 leaves the support points unchanged, but now the weights range, symmetrically of course in $x_1$ and $x_2$, from 0.125 to 0.3125, with the lowest weight on (1,1). When $\chi=1$ we have $\theta_0^2 = \theta_1\theta_2$, so that we are at the lowest value of $\chi$ for which we obtain a design with three support points. All weights are, of course, equal.

These designs were found numerically using a Quasi-Newton algorithm combined with the transformations given in \S9.5 of \citet{ADT:2007}. The general equivalence theorem was used to check the designs by evaluation of the derivative function $\psi(\boldsymbol{x},\xi)$ over a fine grid in ${\cal X}$.  One point in the construction of these designs is that with $\chi = 1$ the optimization algorithm had not quite converged to the theoretical value after the default limit of 100 iterations, whereas around 10 iterations were needed for the other values of $\chi$. The effect on the minimum value of $\psi(\boldsymbol{x},\xi)$ was negligible. A second point is that, to five significant values, the weights for $\chi = 0.5$ were exactly 5/16, 9/32, 9/32 and 1/8. Such simple weights can be an indication that theoretical results are possible. See \citet{aca:2010} and \citet{dette+slava:2011} for an example in discrimination between polynomial regression models.

\begin{figure}[!t]
\centering
{\includegraphics[width=4.2in,clip=true]{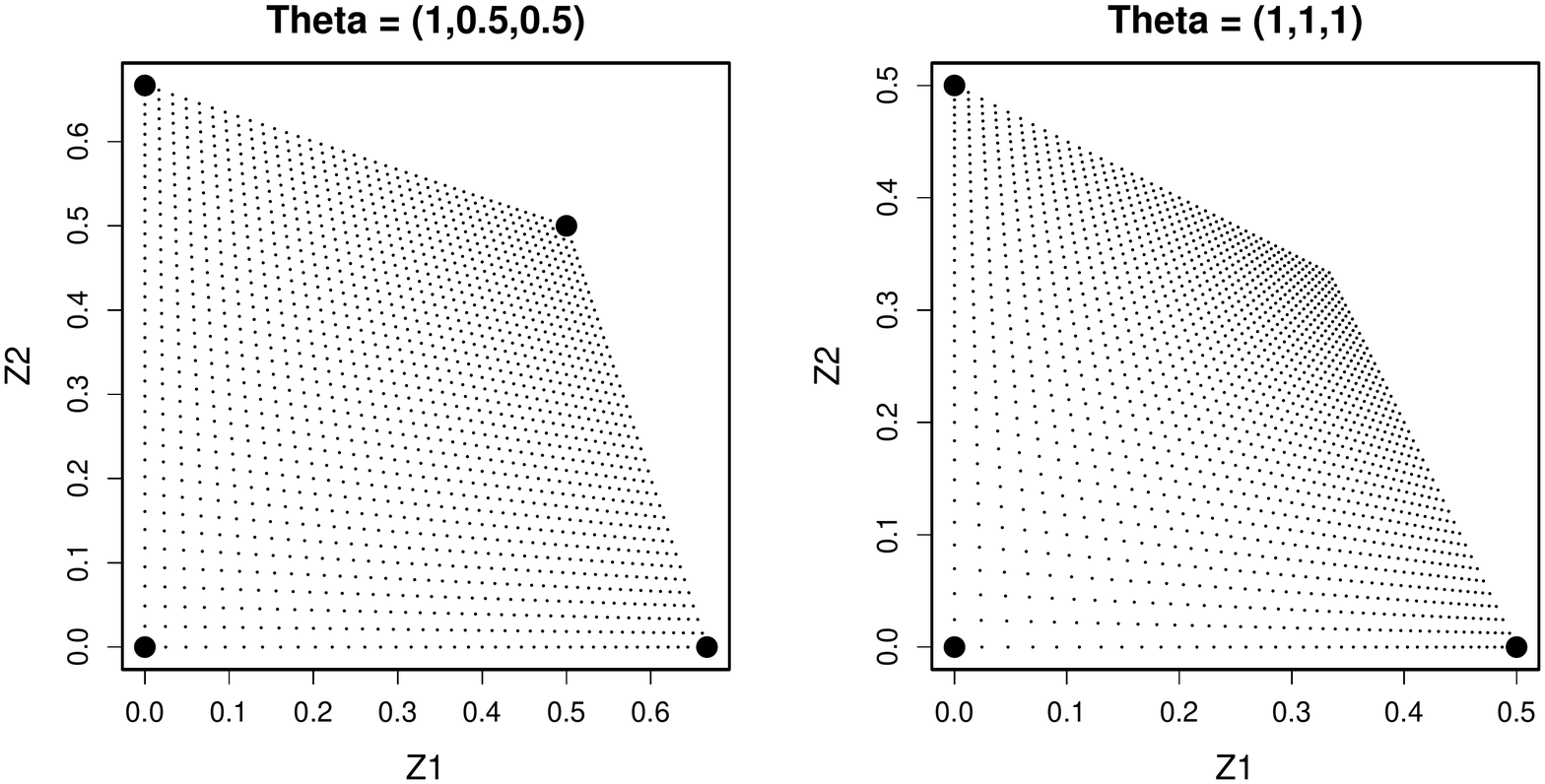}}
\caption{Support points in the induced design region ${\cal Z}$ of D-optimal designs for two-variable first-order model for gamma responses with the power link; $\bullet$ design points \label{glmgamf1}}
\end{figure}

To conclude our discussion of this example, we look at the plots of the design in the induced design region ${\cal Z}$ defined in \eqref{indes2}.   As the results of \citet{burr+s:94} show, the boundary of $\cal Z$ is formed, for $k = 2$, by straight line segments, as is  shown in Figure~\ref{glmgamf1}. There is none of the curving over of the space of $\cal X$ that is caused by the nonlinear nature of the GLM weights $u(\boldsymbol{x}_i)$ for the binomial distribution that is evident in Figure~\ref{fig22b3}.

For small $\chi$ the induced design region is virtually square, becoming less so as $\chi$ increases. The left-hand panel of Figure~\ref{glmgamf1} is for $\chi = 0.5$, for which the weight at $\boldsymbol{x} = (1,1)^\prime$ is 1/8. As $\chi$ increases, the weight on this value decreases. Insight about the case $\chi = 1$ comes from the results of \citet{s+dmt:73} relating D-optimality to minimum volume ellipsoids enclosing design regions. It follows that when $\chi = 1$ the values of $z_1$ and $z_2$ at $\boldsymbol{x} = (1,1)^\prime,$ say $z_1(1,1)$ and $z_2(1,1)$, must be the same distance from the origin as $z_1(1,0)$ and $z_2(1,0)$ (or $z_1(0,1)$ and $z_2(0,1)$). Hence $z_1(1,1) = z_2(1,1)$ = $(\surd 2)/4$. For larger values of $\chi$, the values of $z_1$ and $z_2$ lie inside the circle and $\mathcal{Z}$ becomes increasingly triangular. The design does not change as $\chi$ increases above 1.

In these calculations we have taken $\kappa = 1$. From the form of $u(\boldsymbol{x})$ in \eqref{powerlinku}, other values of $\kappa$ lead to identical figures, but with different numerical values on the axes.

\subsubsection{Second-order response surface with the power link}
\label{respowerlinksec}

\citet[\S6.9]{a+r:2000} use a gamma model to analyse data from \citet{wn:81} on the degradation of insulation due to elevated temperature at a series of times.  A second-order model is required in the two continuous variables and a gamma model fits well with a power link with $\kappa = 0.5$. We scale the variables to obtain design region $\mathcal{X}=[-1,1]^2$. The linear predictor is the quadratic

\begin{equation}
\eta = \theta_0 + \theta_1x_1 + \theta_2x_2 + \theta_{11}x_1^2 +
\theta_{22}x_2^2 + \theta_{12}x_1x_2\,,
\label{22linpred2}
\end{equation}

\noindent that is,~(\ref{22linpred}) with $\gamma = 1$. Then the standard D-optimal design for the normal theory regression model has unequally weighted support at the points of the $3^2$ factorial: weight 0.1458 at the four corners of the design region, 0.0802 at the centre points of the sides and 0.0960 at the centre of the region. This design is optimal for the gamma model with log link and for the model with the power link as the parameters in \eqref{22linpred2}, apart from $\theta_0$, become small. We take $\boldsymbol{\theta}$ to have the values given in Table~\ref{22tabG1}, G1 being rounded from an analysis of Nelson's data.

\begin{table}[!t]
\begin{center}
\caption{Parameter values for linear predictor~\eqref{22linpred2}
with $\kappa = 0.5$ } \label{22tabG1} \vspace{5mm}
\begin{tabular}{c|cccccc} \hline
Parameter set & $\theta_0$ & $\theta_1$  &  $\theta_2$ & $\theta_{11}$ & $\theta_{22}$  & $\theta_{12}$   \\
\hline & & & &
\\
G1     &   $3.7$  & $-0.46$ & $-0.65$ & $-0.19$ & $-0.45$  & $-0.57$ \\
G2 & $3.7$ & $-0.23$ & $-0.325$ & $-0.095$ & $-0.225$ & $-0.285$  \\ & & & &
\\
\hline
\end{tabular}
  \end{center}
  \end{table}

The exact optimal 9-trial design for G1, found by searching over a grid of candidates with steps of 0.01 in $x_1$ and $x_2$, is in Table~\ref{22tabG2}. This shows that, at the points of the design, the minimum value of $\mu$ is 1.90 and the maximum 14.59. The parameter values are thus such that we satisfy the requirement $\mu > 0$.

\begin{table}[!t]
\begin{center}
 \caption{Exact D-optimal designs for the parameter sets G1 and G2 of
Table~\ref{22tabG1}} \label{22tabG2} \vspace{5mm}
\begin{tabular}{r|rrrr||rrrr} \hline
     & \multicolumn{4}{c||}{Design for G1}   & \multicolumn{4}{c}{Design for G2 }     \\
 $i$ & $x_{1i}$ & $x_{2i}$ & $n_i$ & $\mu_i$
     & $x_{1i}$ & $x_{2i}$ & $n_i$ & $\mu_i$  \\
\hline
     &          &          &       &
     &          &          &       &         \\
 1   &   $-1.00$  &   $-1.00$  &   $1$   &  $12.96$
     &   $-1.00$  &   $-1.00$  &   $1$   &  $13.32$  \\
 2   &   $-1.00$  &    $1.00$  &   $2$   &  $11.83$
     &   $-1.00$  &    $1.00$  &   $1$   &  $12.74$  \\
 3   &    $1.00$  &   $-1.00$  &   $2$   &  $14.59$
     &    $1.00$  &   $-1.00$  &   $1$   &  $14.14$  \\
 4   &    $1.00$  &    $1.00$  &   $1$   &   $1.90$
     &    $1.00$  &    $1.00$  &   $1$   &   $6.45$  \\
     &          &          &       &
     &          &          &       &         \\
 5   &    $0.11$  &    $0.15$  &   $1$   &  $12.46$
     &   $-1.00$  &    $0.00$  &   $1$   &  $14.71$  \\
 6   &    $0.26$  &    $1.00$  &   $1$   &   $5.38$
     &   $-0.01$  &  $-1.00$  &   $1$   &  $14.44$  \\
 7   &    $1.00$  &    $0.29$  &   $1$   &   $7.07$
     &    $0.07$  &    $0.09$  &   $1$   &  $13.33$  \\
 8   &          &          &       &
     &    $0.08$  &    $1.00$  &   $1$   &   $9.66$  \\
 9   &          &          &       &
     &   $1.00$   &    $0.09$  &   $1$   &  $11.01$  \\
     &          &          &       &
     &          &          &       &         \\
\hline
\end{tabular}
  \end{center}
  \end{table}


\begin{figure}[t]
   \begin{center}
   \includegraphics[width=4.0in,clip=true]{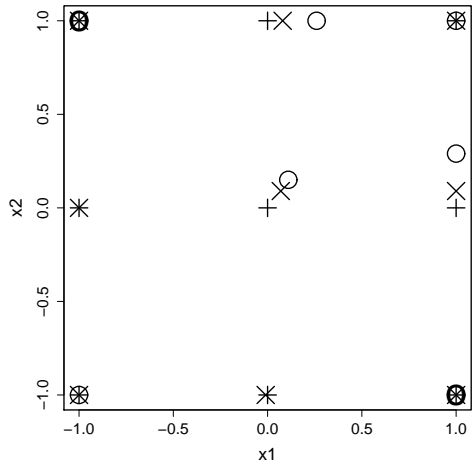} \vspace{-5mm}
   \caption{Points for D-optimal 9-point designs for gamma models in
   Table~\ref{22tabG2}: $+$, the points of the $3^2$ factorial;
   $\circ$, G1 and $\times$, G2.  Points for G1 which are replicated
   twice are darker.}
   \label{fig22g1}
   \end{center}
\end{figure}

As the left-hand half of Table~\ref{22tabG2} shows, the design has seven support points. The points of the $2^2$ factorial are in the upper part of the table. All are included in the design, two being replicated. The other three support points are slight distortions of some remaining points of the support of the $3^2$ factorial. Figure~\ref{fig22g1} makes clear the virtually symmetrical nature of the design, although the parameters are not quite symmetrical in value for $x_1$ and $x_2$.

To illustrate the approach of the design to the $3^2$ factorial as the parameter values decrease we also found the D-optimal 9-point design for the set of parameter values G2 in Table~\ref{22tabG1} in which all parameters, other than $\theta_0$, have half the values they have for design G1. As Table~\ref{22tabG2} shows, the range of means at the design points is now 6.45 to 14.71, an appreciable reduction in the ratio of
largest to smallest response. The support points of the design for G2 are shown in Figure~\ref{fig22g1} by the symbol \textsf{X}.  There are now nine distinct support points close to those of the $3^2$ factorial. For G2 the three design points in the lower half of the Table for G1 are moved in the direction of the full factorial design. For linear regression the unweighted $3^2$ factorial is D-optimal.

\subsubsection{Efficient standard designs for gamma models}
\label{22gamsdsec}

We conclude our analysis of designs for the gamma model with second-order predictor and a power link by briefly exploring how well the unweighted $3^2$ factorial performs  when there are nine trials by comparing both it and the design for G2 with that for G1.

The D-optimal design for the less extreme parameter set G2 of Table~\ref{22tabG1} has efficiency 97.32\%, while the equi-replicated $3^2$ factorial has efficiency 96.35\%. The main feature of these designs is how efficient they are for the gamma model, both efficiencies being greater than 95\%. The design for parameters G2 is for a model with smaller effects than G1, so that the design and its efficiency are between those for G1 and the factorial design.

An indication of this example with a gamma response is that standard designs may be satisfactory for  second-order response surfaces. However, \cite{burr+s:94} show that, for first-order models, full $2^2$ factorial designs, or their regular fractions (see Chapter~7), can be very inefficient when the effects are strong and the optimal designs have only $k+1$ points of support.

\subsubsection{Designs for the power link and for the Box-Cox link}
\label{twolinkssec}

We now briefly explore the relationship between designs for the power link with weights given by~\eqref{powerlinku} and those for the Box-Cox link~\eqref{eqg3}. We relate the two through their dependence on the linear predictor $\eta$.

From  \eqref{powerlinku} the weights for the power link can be written

\begin{equation}
\{u(\boldsymbol{x})\}^{-0.5} = \eta\,,
\label{powerusqrt}
\end{equation}

\noindent since the constant value of $\kappa$ is ignored. For the Box-Cox link, on the other hand,

\begin{equation}
\{u(\boldsymbol{x})\}^{-0.5} = \mu^{\lambda}\,.
\label{bcusqrt}
\end{equation}

\noindent However, from \eqref{eqg1}

\[ \mu = (1 + \lambda \eta)^{1/\lambda}\,, \]

\noindent so that, for the Box-Cox link,
\[
\{u(\boldsymbol{x})\}^{-0.5}  = (1 + \lambda \eta) = 1 + \lambda(\theta_0 + \theta_1 x_1 +\ldots+\theta_k x_k)\,.
\]

\noindent The condition in Theorem~2 of \S\ref{powerlinksec} that the one-factor-at-a-time design is optimal therefore becomes

\begin{equation}
(1 + \lambda \theta_0)^2 \le \lambda^2\theta_i\theta_j\,.
\label{bcusqrt2}
\end{equation}

An advantage of the Box-Cox link in data analysis is that it is continuous at $\lambda =0$, becoming the log link. The search over suitable links to describe the data therefore does not contain any discontinuity. In designing experiments, on the other hand, a locally optimal design will be selected for a particular $\lambda$. The results of \S\ref{gamsec} show that, if the power link is used, a value of $\kappa$ does not have to be stated \textit{a priori}. However, prior values will be required for $\boldsymbol{\theta}$. These will typically be ascertained from guessed responses as the factors vary. Despite the absence of explicit reference to $\kappa$ in the design criterion, the value will enter implicitly through the relationship between $\mu$ and $\eta$ \eqref{powerlink}. Finally, \eqref{bcusqrt2} shows that, as $\lambda \rightarrow 0$, the one-factor-at-a-time design will not be optimal. Further, from  \eqref{bcusqrt}, it follows that, under these conditions the weights $u(\boldsymbol{x}) \rightarrow 1$ and  the design will tend to the $2^k$ factorial, even for large values of the $\theta_j$.

\subsection{Poisson data}
\label{poissondesigns}

D-optimal designs for Poisson regression with the log link, $\log\mu=\eta$, share some similarities with the Gamma designs in \S\ref{gamdatsec}. In particular, for log linear models with a first-order linear predictor,

\begin{equation}\label{loglin}
\log\mu = \eta  = \theta_0 + \sum_{i=1}^{k}\theta_ix_{i}\,,
\end{equation}

\noindent the optimal design has a similar structure to those from Theorem~2.

There is only a moderate number of results on designs for Poisson regression in the literature. For~\eqref{loglin} and $k=1$, \citet{mink:93} found locally optimal designs for estimating $\theta_1$; see also Chapter~14 and the references therein for more general results on models with one variable. For $k=1,2$, \citet*{wmsy} investigated the dependence of locally optimal designs on functions of the parameter values and \citet*{wsy} developed sequential designs. For a single variable, \citet{ftw:92} used a transformation of the design space to a canonical form, together with geometrical arguments along the lines of \S\ref{inducesec}, to find locally optimal designs for a class of nonlinear models that included Poisson regression.

\begin{figure}[t]
\begin{center}
\includegraphics[scale=0.6]{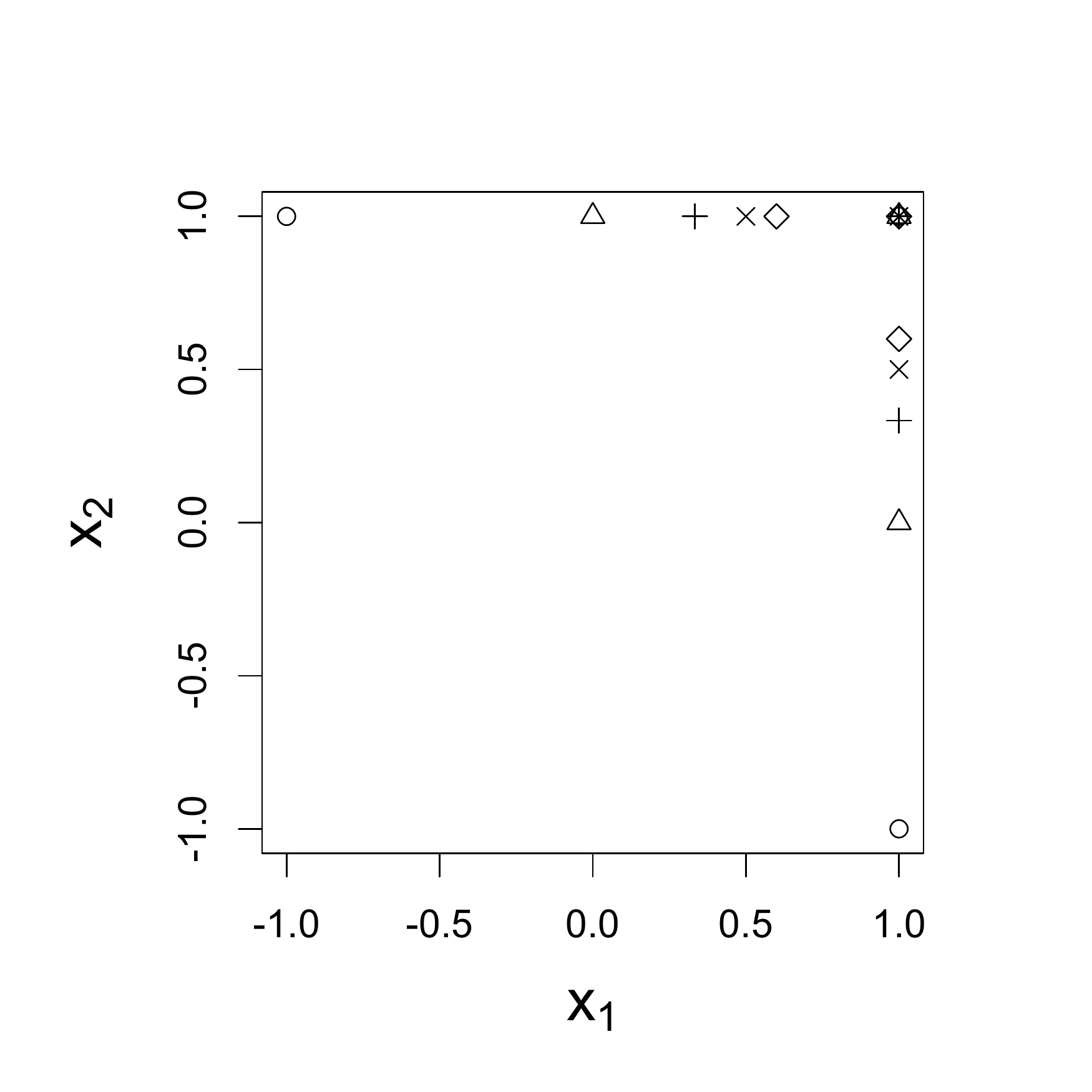}
\end{center}
\caption{\label{poisex}Support points for locally D-optimal designs for Poisson regression with~\eqref{loglin} and $\boldsymbol{\theta}= (0,\chi,\chi)$. Key: $\circ$ $\chi=1$; $\triangle$ $\chi=2$; $+$ $\chi=3$; $\times$ $\chi=4$; $\diamond$ $\chi=5$. All designs include the point $(1,1)$.}
\end{figure}

\citet{russpoiss:2009} addressed the problem of D-optimal design for~\eqref{loglin} with $k\ge 1$ and provided the following theorem.

\vspace{0.25cm}

\textbf{Theorem~3}. \citep{russpoiss:2009} \textit{A D-optimal design for Poisson regression~\eqref{loglin} with $l_i\le x_{ij}\le u_i$ and $|\theta_i(u_i-l_i)|\ge 2$ $(i=1,\ldots,k; j=1,\ldots,t)$ has the $t=k+1$ equally weighted support points}

\begin{eqnarray}
\boldsymbol{x}_i & = & \boldsymbol{c} - \frac{2}{\theta_i}\boldsymbol{e}_i\,,\quad i=1,\ldots,k \label{poissopt}\\
\boldsymbol{x}_{k+1} & = & \boldsymbol{c}\,,\nonumber
\end{eqnarray}

\noindent \textit{for $\boldsymbol{e}_i$ the $i$th column vector of the $k\times k$ identity matrix, $i=1,\ldots,k$, and $\boldsymbol{c} = (c_1,\ldots,c_k)^\prime$, where $c_i=u_i$ if $\theta_i>0$ and $c_i=l_i$ if $\theta_i<0$.}

\vspace{0.25cm}

\begin{figure}[t]
\begin{center}
\includegraphics[scale=0.65]{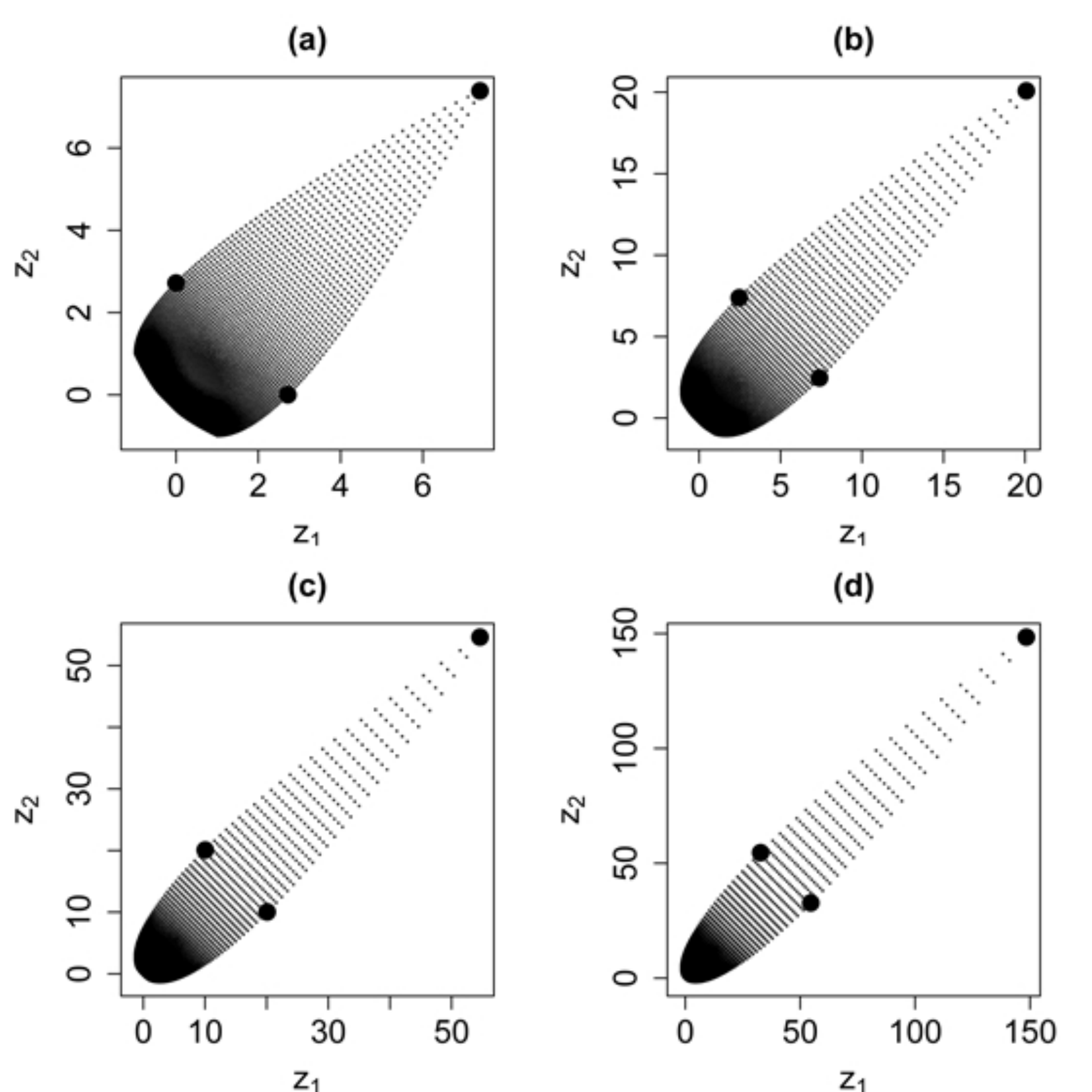}
\end{center}
\caption{\label{zpois}Support points in the induced design space $\mathcal{Z}$ for D-optimal designs for Poisson regression with~\eqref{loglin} and $\boldsymbol{\theta}= (0,\chi,\chi)$. (a) $\chi=2$; (b) $\chi=3$; (c) $\chi=4$; (d) $\chi=5$.}
\end{figure}

The proof of Theorem~3 is via a canonical transformation and an application of the general equivalence theorem. Note that the D-optimal design does not depend on the value of the intercept, $\theta_0$, and is invariant to permutation of the factor labels. The requirement $|\theta_i(u_i-l_i)|\ge 2$ is not overly restrictive in practice; $\mathcal{X}=[-1,1]^k$ requires $|\theta_i|\ge 1$, $i=1,\ldots,k$. In Figure~\ref{poisex}, we give the support points for $k=2$ and a number of example parameter vectors, $\boldsymbol{\theta}=(0,\chi,\chi)^\prime$. Notice the one-factor-at-a-time structure of the design and how the support points tend towards $(1,1)$ as $\chi$ increases. Figure~\ref{zpois} gives the support points in the induced design space $\mathcal{Z}$, projected into $z_1,z_2$ and defined from equation~(\ref{indes2}). The optimal support points lie on the boundary of the induced space. Not only do the values of the $z_i$ increase with $\chi$, but the induced design region itself becomes more elongated as $\chi$ increases.

For $|\theta_i(u_i-l_i)| < 2$, numerical investigation has found that both the optimal support points and weights depend on $\theta_0$, in addition to the other parameters. As expected, as $|\theta_i/\theta_0|$ tends to zero, for $i=1,\ldots,k$, the D-optimal design tends to the equally weighted factorial design.

For more general linear predictors, for example containing interactions or quadratic terms, numerical search is required to find optimal designs. This is also the case for Bayesian D-optimal designs except for the special case of minimally-supported designs for~\eqref{loglin}, that is, designs with $t=k+1$ support points. \citet{me12} provided theoretical results for minimally-supported designs robust to a set of models of the form~\eqref{loglin} defined through a discrete set of parameter vectors. We extend their result to Bayesian minimally supported D-optimal designs.

\vspace{0.25cm}

\textbf{Theorem~4}: \textit{Assume a Poisson regression model with linear predictor~\eqref{loglin}. The Bayesian D-optimal design among the class of minimally supported designs, minimising~\eqref{baysdcrit}, is the locally D-optimal design~\eqref{poissopt} for the parameter vector $\boldsymbol{\theta}^\star = E(\boldsymbol{\theta})$ provided $|E(\theta_i)(u_i-l_i)|\ge 2$.}

\textbf{Proof}: \textit{For a minimally-supported design, the model matrix}

$$\boldsymbol{X}=[\boldsymbol{f}(\boldsymbol{x}_1),\ldots,\boldsymbol{f}(\boldsymbol{x}_p)]^\prime$$

\noindent \textit{is $p\times p$.  Now the objective function~\eqref{baysdcrit} can be written as}

\begin{eqnarray}
\Psi(\xi) & = & -\int_\Theta \log |\boldsymbol{M}(\xi;\,\boldsymbol{\theta})|p(\boldsymbol{\theta})d\boldsymbol{\theta} \label{poisdopt}\\
 & = &  -\int_\Theta 2\log |\boldsymbol{X}|p(\boldsymbol{\theta})d\boldsymbol{\theta} -\int_\Theta \log \prod_{j=1}^tw_j\exp(\eta_j)p(\boldsymbol{\theta})d\boldsymbol{\theta}\nonumber\\
 & = & -2\log |\boldsymbol{X}| -\sum_{j=1}^t\left[\log w_j +\int_\Theta \eta_jp(\boldsymbol{\theta})d\boldsymbol{\theta}\right]\nonumber\\
 & = & -2\log |\boldsymbol{X}| -\sum_{j=1}^t\left[\log w_j -\eta_j^\star\right]\nonumber\\
  & = & -\log |\boldsymbol{M}(\xi;\, \boldsymbol{\theta}^\star)|\,, \label{bayeslocopt}
\end{eqnarray}

\noindent \textit{where $\boldsymbol{\theta}^\star = E(\boldsymbol{\theta})$ and $\eta_j^\star = \theta_0^\star + \sum_{i=1}^k\theta_i^\star x_{ij}$. The fourth line above follows as $\eta_i$ is a linear function of $\boldsymbol{\theta}$. The equality of~\eqref{poisdopt} and~\eqref{bayeslocopt} establishes that, provided $|E(\theta_i)(u_i-l_i)|\ge 2$, design~\eqref{poissopt} is Bayesian D-optimal among the class of minimally supported designs.}

\vspace{0.25cm}

To illustrate Theorem~4, we find a Bayesian minimally supported D-optimal design for~\eqref{loglin} and $k=5$ factors with $x_i\in [-1,1]$, $\theta_0=0$ and each $\theta_i\sim U(a,b)$ $(i=1,\ldots,5)$. The values of $a$ and $b$ are given in Table~\ref{poisrobustdes} (a) in terms of a common parameter $\alpha$. Increasing $\alpha$ leads to more diffuse prior densities. However, for any $\alpha\ge 2$, the Bayesian minimally-supported D-optimal design is given by the locally D-optimal design for $\theta_0=0$ and $\theta_i= (a+b)/2=(-1)^{(i+1)}(1+\alpha/2)$; see Table~\ref{poisrobustdes} (b).

\begin{table}[h]
\caption{ \label{poisrobustdes}Bayesian minimally supported D-optimal design: (a) Ranges for the uniform, $U(a,b)$, prior densities for $\theta_1,\ldots,\theta_5$; (b) Equally weighted support points; $\beta=\left[(\alpha-2)/(\alpha+2)\right]$ for $\alpha=2,5,10,20$.}
\begin{center}
\begin{minipage}{6cm}
\centerline{(a) Parameter ranges}
$$
\begin{array}{crr}
\hline
 & \multicolumn{2}{c}{\mbox{Limits}} \\
\mbox{Parameter} & a & b \\
\hline
\theta_1 & 1 & 1+\alpha\\
\theta_2 & -1-\alpha & -1\\
\theta_3 & 1 & 1+\alpha\\
\theta_4 & -1-\alpha & -1\\
\theta_5 & 1 & 1+\alpha\\
\hline
\end{array}
$$
\end{minipage}
\begin{minipage}{6cm}
\centerline{(b) Support points}
$$\begin{array}{crrrrr}
\hline
 & x_1 & x_2 & x_3 & x_4 & x_5\\
\hline
1 & \beta & -1 & 1 & -1 & 1 \\
2 & 1 & -\beta & 1 & -1 & 1 \\
3 & 1 & -1 & \beta & -1 & 1 \\
4 & 1 & -1 & 1 & -\beta & 1 \\
5 & 1 & -1 & 1 & -1 & \beta \\
6 & 1 & -1 & 1 & -1 & 1 \\
\hline
\end{array}$$
\end{minipage}
\end{center}
\normalsize
\end{table}

\begin{figure}[t]
\begin{center}
\includegraphics[scale=0.65]{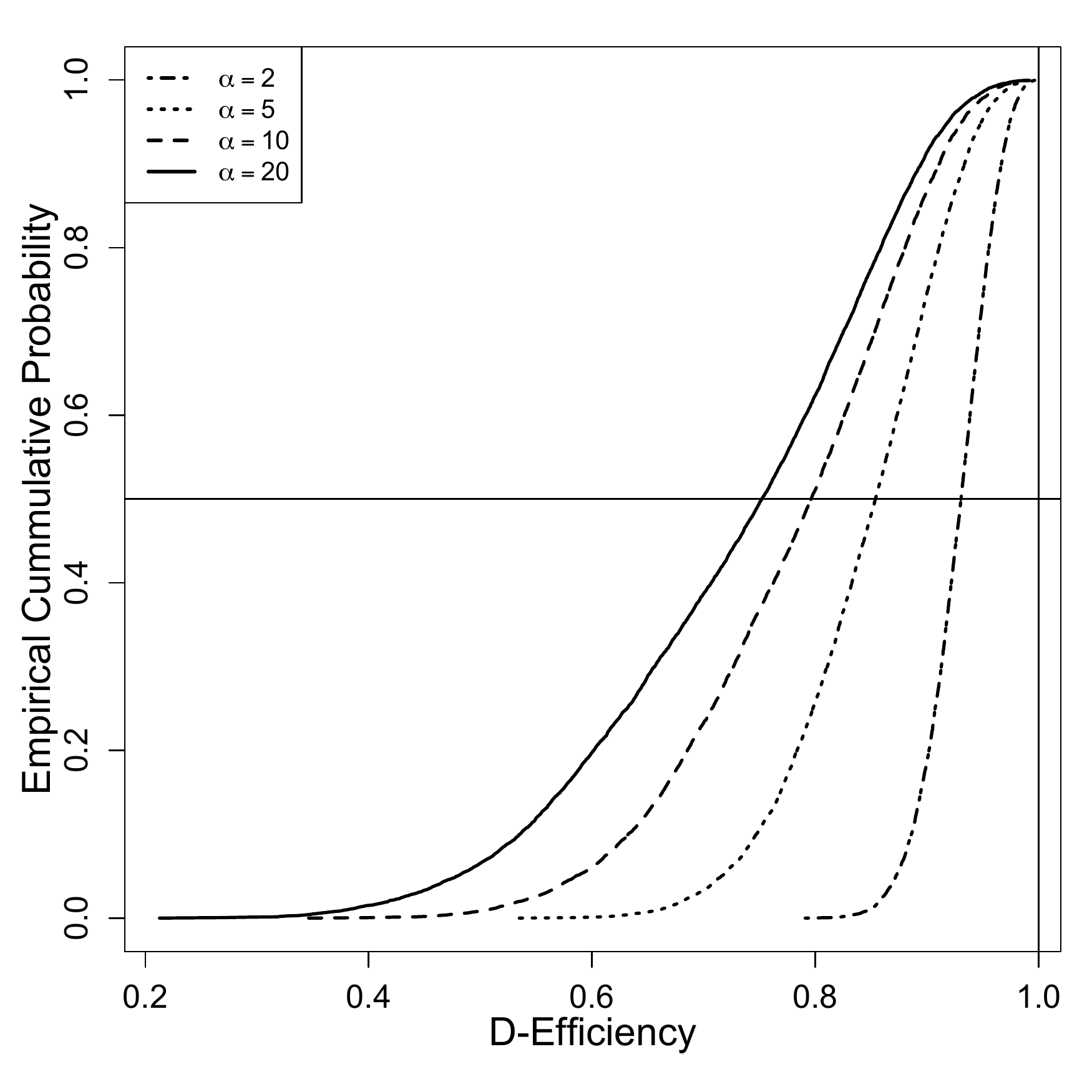}
\end{center}
\caption{\label{poissrobust} Empirical cumulative distributions for the D-efficiency of the Bayesian minimally supported D-optimal design for four different prior distributions;  see Table~\ref{poisrobustdes}.}
\end{figure}

We assess the performance of these designs through simulation of 10,000 parameter vectors from the uniform distributions defined by Table~\ref{poisrobustdes} (a) for  $\alpha=2,5,10$ and $20$. For each parameter vector, we derive the locally D-optimal design from Theorem~3, and then calculate the D-efficiency~\eqref{deffic} for the design in Table~\ref{poisrobustdes} (b). The induced empirical cumulative distribution functions are given in Figure~\ref{poissrobust}.

For relatively precise prior information ($\alpha=2$), high efficiency is maintained for all the samples from the prior distribution, with minimum efficiency of 79\% and median of 93\%. As the prior distribution becomes more diffuse (with increasing $\alpha$), the induced efficiency distribution also becomes more diffuse. For $\alpha=5$, the minimum and median efficiencies are 53\% and 85\% respectively; the corresponding summaries for $\alpha=10$ are 34\% and 80\% and for $\alpha=20$ are 21\% and 75\%. The high median efficiencies maintained for more diffuse distributions are not typical of minimally supported designs in general (for example, see \citealp{vdvw2014} for binary data designs). However, it seems the structure of the Poisson designs, with all the support points on the boundary, makes Bayesian minimally supported designs an efficient choice (see also \citealp{me12})

\section{Designs with dependent data}
\label{corr}

There is a variety of practical experiments in which the observed responses may be dependent; see also Chapter~5. Most obviously, and perhaps most importantly, are experiments where there is a natural grouping in the experimental units such that, given the same treatment, two observations from within the same group are expected to be more similar than two observations in different groups. Common examples for non-normal data include longitudinal studies in clinical and pharmaceutical trials \citep{dhlz2002} and blocked experiments in industry \citep{rmm04}. Our exposition in this section focuses on block designs but is equally appropriate for other grouped experiments.

For experiments involving the observation of a continuous response that is appropriately described by a normal distribution, there is, of course, a wealth of literature on the design and analysis of blocked experiments for linear models (see Part II of this book). There is also a considerable literature on design for nonlinear models and dependent data, much of it stemming from the world of pharmacokinetics/phamacodynamics (PK/PD) and the seminal work of \citet{mmb:97}.  For experiments with discrete data, with a binary or count response, there are rather fewer results available, although the results we describe below naturally share some similarities with the PK/PD literature, particularly the so-called first-order approximation (see \citealp{rm2003} and \citealp{brm2010}).

As with linear models, the first decision is whether to model the block effects as fixed or random. We will choose the latter option and take a mixed model approach as: (i) in industrial experiments, the blocks are usually a nuisance factor and not of interest in themselves; (ii) random block effects allow prediction of the response for unobserved blocks; and (iii) pragmatically, when finding optimal designs for nonlinear models, the use of random block effects reduces the number of unknown parameters for which prior information is required. For example, with $b$ blocks, a fixed-effects design would require the specification of $b-1$ block effects. See \citet{sy2012} for locally optimal designs with fixed group effects and a single variable.

\subsection{Random intercept model}
\label{glmm}

To model the responses from a blocked experiment, we adopt the framework of Generalized Linear Mixed Models (GLMMs; \citealp{bc}) and, in particular, apply random intercept models. We develop our framework for $b$ blocks each of equal size $m$. For the $j$th unit in the $i$th block,

$$
y_{ij}| \gamma_i \sim \pi(\mu_{ij}) \,, \qquad \mbox{for } i=1,\ldots,b;\, j=1,\ldots,m\,,
$$

\noindent where $\pi(\cdot)$ is a distribution from the exponential family with mean $\mu_{ij}$ and

$$g(\mu_{ij}) = \boldsymbol{f}^\prime(\boldsymbol{x}_{ij}) \boldsymbol{\theta} + \gamma_i.$$

\noindent Here $g(\cdot)$ is the link function and $\boldsymbol{x}_{ij}$ is the $i,j$th combination of variable values. As before, the vector $\boldsymbol{f}(\boldsymbol{x}_{ij})$ holds known functions of the $k$ variables and $\boldsymbol{\theta}$ holds the $p$ unknown regression parameters. The unobservable random block effects $\gamma_i$ are assume to follow independent $N(0,\sigma_\gamma^2)$ distributions, with $\sigma_\gamma^2$ known.  Under this model, observations in different blocks are independent. More general models including additional random effects, such as random slopes, may also be defined.

\subsection{Continuous block designs}
\label{contblock}

We choose to generalise~(\ref{optdes}) to include blocks of fixed size $m$ through

$$
\xi = \left\{
    \begin{array}{ccc}
        \zeta_1& \ldots& \zeta_t \\
        w_1 &\ldots& w_t
    \end{array}
\right\}\,,
$$

\noindent where $\zeta_{l} \in \mathcal{X}^m$ is the set of design points that form the $l$th block (or ``support point"), $0 < w_l \leq 1$ is the corresponding weight; $\sum_{l=1}^{t} w_l = 1$. See \citet{Cheng1995} and \citet{wv}. For example, if $k=m=2$, a possible continuous design is

$$
\xi = \left\{
    \begin{array}{cc}
    \zeta_1 = \{(-1,-1), (1,1)\} & \zeta_2=\{(1,-1), (-1,1)\} \\
    0.5 & 0.5
    \end{array}
\right\}\,,
$$

\noindent that is, one-half of the $b$ blocks in a realised exact design would contain design points $x_1=x_2=-1$ and $x_1=x_2=1$, and the other half would contain design points $x_1=1,x_2=-1$ and $x_1=-1,x_2=1$.

\subsection{Information matrix for a random intercept model}

To apply the usual design selection criteria, for example, D-optimality, we need to derive and evaluate the information matrix for $\boldsymbol{\theta}$. As observations in different blocks are independent,

$$
\boldsymbol{M}(\xi;\,\boldsymbol{\tau}) = \sum_{l=1}^{t} w_l \, \boldsymbol{M}(\zeta_l;\,\boldsymbol{\tau})\,,
$$

\noindent where $\boldsymbol{\tau} = (\boldsymbol{\theta}^\prime, \sigma_\gamma^2)$, and the information matrix for the $l$th block is, by definition,

\begin{equation}\label{blockinfo}
\boldsymbol{M}(\zeta_l, \boldsymbol{\tau}) = E_{\boldsymbol{y}_l}  \left\{
                                            - \frac{\partial^2 \log p( \boldsymbol{y}_l | \boldsymbol{\tau},\zeta_l)}
                                                    {\partial \boldsymbol{\theta} \partial \boldsymbol{\theta}^\prime}
                                        \right\}\,,
\end{equation}

\noindent where the $m$-vector $\boldsymbol{y}_l=(y_{l1},\ldots.y_{lm})^\prime$ holds the responses from block $\zeta_l$.

%
%

Direct calculation of the expectation in~(\ref{blockinfo}) is possible for small experiments with binary data through

$$
\boldsymbol{M}(\zeta_l, \boldsymbol{\tau}) = \sum_{ \boldsymbol{y}_l \in \{0,1\}^m }
        - \frac{\partial^2 \log p( \boldsymbol{y}_l | \boldsymbol{\tau},\zeta_l)}
                                                    {\partial \boldsymbol{\theta} \partial \boldsymbol{\theta}^\mathrm{T}}
                        p( \boldsymbol{y}_l | \boldsymbol{\tau},\zeta_l)\,,
$$

\noindent although both the marginal likelihood and its derivative will require numerical approximation \citep[see][]{waite}. For more practically-sized experiments, this direct calculation will be computationally infeasible.

\subsection{Approximating the information matrix using estimating equations}
\label{glmmapprox}

For model estimation, the marginal likelihood can be approximated using methods such as quadrature, Markov Chain Monte Carlo or the EM algorithm \citep{msn}. However, for the purposes of finding an optimal design using numerical search, repeated evaluation of the information matrix, or some alternative, is required. Hence, a fast approximation to~(\ref{blockinfo}) is needed.

An approximate variance-covariance matrix for $\boldsymbol{\theta}$ is available from the theory of estimating equations (see, for example, \citealp{godambe}). For a GLMM, standard unbiased estimating equations are an extension of the score equations for a GLM and have the form

$$
\sum_{l=1}^tw_l\boldsymbol{X}_l^\prime\boldsymbol{\Delta}_l\boldsymbol{V}^{-1}_l(\boldsymbol{y}_l - \boldsymbol{\mu}_l) = \mathbf{0}\,,
$$

\noindent where $\boldsymbol{\Delta}_l = \mathrm{diag}\left[d\mu_{lj}/d\eta_{lj}\right]$, $\eta_{lj} = \boldsymbol{f}^\prime(\boldsymbol{x}_{lj})\boldsymbol{\theta}$, and $\boldsymbol{X}_l$ and $\boldsymbol{\mu}_l$ are the $m\times p$ model matrix and $m\times 1$ mean vector defined for the $l$th block; $\boldsymbol{V}_l$ is a $m\times m$ weight matrix for the observations from the $l$th block.  Depending on the approximation, $\boldsymbol{\mu}_l$ may be either the \textit{marginal} mean response or the \textit{conditional} mean response given $\gamma_l=0$.

The approximate variance-covariance of the resulting estimators is given by

\begin{equation}\label{asymvar}
\mbox{Var}\left(\hat{\boldsymbol{\theta}}\right) \approx \left(\sum_{l=1}^tw_l\boldsymbol{X}_l^\prime\boldsymbol{\Delta}_l\boldsymbol{V}_l^{-1}\boldsymbol{\Delta}_l\boldsymbol{X}_l\right)^{-1}
\end{equation}

\noindent The inverse of this variance-covariance matrix can be used as an approximation to the information matrix $\boldsymbol{M}(\xi,\boldsymbol{\tau})$.

Various different choices of $\boldsymbol{V}_l$ and $\boldsymbol{\mu}_l$ have been proposed, of which we will discuss the following three:

\begin{enumerate}

\item Quasi-Likelihood (QL): $\boldsymbol{V}_l = \mbox{Var}\left(\boldsymbol{Y}_l\right)$ and $\boldsymbol{\mu}_l=E(\boldsymbol{Y}_l)$, the marginal variance and mean; see \citet{wedderburn}. The marginal variance-covariance matrix for $\boldsymbol{Y}_l$ is generally not available in closed-form for non-normal data. One notable exception is for the Poisson distribution, for which \citet{niaparast}  and \citet{ns} used quasi-likelihood to find D-optimal designs.

\item Marginal quasi-likelihood (MQL): $\boldsymbol{V}_l = \mathrm{diag}\left[\mbox{Var}(Y_{lj})\right] + \boldsymbol{\Delta}_l\boldsymbol{J}\boldsymbol{\Delta}_l\sigma_\gamma^2$, with $\boldsymbol{J} = \boldsymbol{1}\boldsymbol{1}^\prime$, and $\boldsymbol{\mu}_l = E(\boldsymbol{Y}_l | \gamma_l=0)$; see \citet{bc}. Here, a linear mixed model approximation is used for the marginal variance-covariance of $\boldsymbol{Y}_l$. This approximation has been used to find designs for binary data by authors such as \citet{mm}, \citet{ttb} and \citet{oa}.

\item Generalized estimating equations (GEEs):

$$
\boldsymbol{V}_l = \left\{\mathrm{diag}\left[\mbox{Var}(Y_{lj})\right]\right\}^{1/2} \boldsymbol{R}\left\{\mathrm{diag}\left[\mbox{Var}(Y_{lj})\right]\right\}^{1/2}\,,$$

\noindent with $\boldsymbol{R}$ an intra-block marginal ``working correlation'' matrix, and $\boldsymbol{\mu}_l = E(\boldsymbol{Y}_l | \gamma_l=0)$; see \citet{lz86}. The matrix $\boldsymbol{R}$ is assumed to be independent of $\boldsymbol{x}$ and usually chosen to have a standard form, for example, exchangeable or known up to a small number of correlation parameters. For discrete data, it is in fact often impossible for either of these two assumptions to hold but the resulting estimators can be relatively efficient compared to a full maximum likelihood approach \citep{cj}. Methodology for D-optimal designs using this approach was developed by \citet{wv}.

\end{enumerate}

Note that for each of these approximations, if $\sigma_\gamma^2\rightarrow 0$ (or equivalently, $\boldsymbol{R}\rightarrow \boldsymbol{I}$), the variance-covariance matrix for $\boldsymbol{Y}_l$ reverts to that for a simple GLM. \citet{www} developed, assessed and compared a variety of methods of approximating $\boldsymbol{M}$ to find D-optimal designs for a GLMM.

%


\subsection{Comparison of approximations}
\label{glmmcomp}

We use a small example to perform a simple comparison of designs from the three approximations in the previous section. Consider an experiment in blocks of size $m=2$ with a single variable $x$ to collect count data. Conditional on the random block effect, we assume $Y_{ij}|\gamma_i \sim \mbox{Poisson}(\mu_{ij})$ $(i=1,\ldots,b;\,j=1,\ldots,m)$ and we choose a second-order predictor and the log link

$$
\log(\mu_{ij}) = \eta_{ij} = \gamma_i + \theta_0+ \theta_1x_{ij} + \theta_2x_{ij}^2\,,
$$

\noindent for $x_{ij}\in[-1,1]$. For the purposes of finding designs, we assume point prior information and set $\theta_0=0$, $\theta_1=5$ and $\theta_2=1$. The random block effect has distribution $\gamma_i\sim N(0,\sigma_\gamma^2)$ for $i=1,\ldots,b$ and $\sigma_\gamma^2=0.5$.

For the log link, $\boldsymbol{\Delta}_l = \mbox{diag}\left\{\mu_{lj}\right\}$. We consider each of the three approximations, with $\mu_{lj}$ representing either the marginal or conditional mean response for the $j$th point in the $l$th block of support $(l=1,\ldots,t\,;j=1,\ldots,m)$.

\begin{table}[b]
\caption{\label{compdes}D-optimal continuous block designs with blocks of size $m=2$ for a Poisson example and Quasi-Likelihood, Marginal Quasi-Likelihood and Generalized Estimating Equations approaches (to 2 decimal places).}
   \begin{center}
\begin{tabular}{cccc}
\hline
 &  \multicolumn{3}{c}{Support blocks} \\
 & Block 1 & Block 2 & Block 3 \\
\hline
QL/MQL & (0.88,0.10) & (1,0.75) & -- \\
Weights & 0.5 & 0.5 & -- \\
GEE & (0.84,0.02) & (0.72, 1) & (1,0.26) \\
Weights & 0.38 & 0.35 & 0.27 \\
\hline
\end{tabular}
\end{center}
 \end{table}

\vspace{0.25cm}

\noindent \textit{Quasi-Likelihood}: Here, $\mu_{lj} = E(Y_{lj}) = \exp\left(\eta_{lj}+\sigma_\gamma^2/2\right)$, the marginal mean response, and

$$
\boldsymbol{V}_l = \mbox{diag}\left\{\exp\left(\eta_{lj}+\sigma_\gamma^2/2\right)\right\} + \left\{\exp\left(\sigma_\gamma^2\right)-1\right\}\bar{\boldsymbol{\mu}}_l\bar{\boldsymbol{\mu}}_l^\prime\,,
$$

\noindent where $\bar{\boldsymbol{\mu}}_l^\prime = \left\{\exp\left(\eta_{lj}+\sigma_\gamma^2/2\right)\right\}_{j=1}^m$.

\vspace{0.25cm}

\noindent \textit{Marginal Quasi-Likelihood}: For this approximation, $\mu_{lj} = \exp\left(\eta_{lj}\right)$, the conditional mean response given $\gamma_l=0$, and

$$
\boldsymbol{V}_l = \mbox{diag}\left\{\exp\left(\eta_{lj}\right)\right\} + \sigma_\gamma^2\boldsymbol{\mu}_l\boldsymbol{\mu}_l^\prime\,,
$$

\noindent where $\boldsymbol{\mu}_l^\prime = \left\{\exp\left(\eta_{lj}\right)\right\}_{j=1}^m$.

\vspace{0.25cm}

\noindent \textit{Generalized Estimating Equations}: Now, $\mu_{lj} = \exp\left(\eta_{lj}\right)$, the conditional mean response given $\gamma_l=0$, and

$$
\boldsymbol{V}_l = \mbox{diag}\left\{\exp\left(\eta_{lj}\right)\right\}^{\frac{1}{2}}\boldsymbol{R}\mbox{diag}\left\{\exp\left(\eta_{lj}\right)\right\}^{\frac{1}{2}}\,,
$$

\noindent where the working correlation matrix for this example is

$$
\boldsymbol{R} =
\left(
\begin{array}{cc}
1 & \alpha \\
\alpha & 1
\end{array}
\right)\,.
$$

\noindent For the GEE design, we redefine $\boldsymbol{\tau}=(\boldsymbol{\theta}^\prime,\alpha)$.

Locally D-optimal designs under the three approximations are given in Table~\ref{compdes}. Note that the same design was found under the QL and MQL approximations and that, for the GEE design, $\alpha=0.5$ was chosen so that the working correlation closely matched the average intra-block correlation ($\approx 0.49$) from the other design. Table~\ref{blockeff} gives the D-efficiencies~\eqref{deffic} of each design under each approximation; the GEE design is 87\% efficient under the QL and MQL approximations, whilst the QL/MQL design is 90\% efficient under the GEE approximation.

\begin{table}
\caption{\label{blockeff}Efficiencies of three optimal designs under three approximations to the information matrix.}
\begin{center}
\begin{tabular}{ccc}
 & \multicolumn{2}{c}{Approximation} \\
Design & QL/MQL & GEE \\
\hline
QL/MQL & 1 & 0.90 \\
GEE & 0.87 & 1 \\
\hline
\end{tabular}
\end{center}
\end{table}

\begin{figure}
\begin{center}
\begin{tabular}{cc}
\includegraphics[scale=0.375]{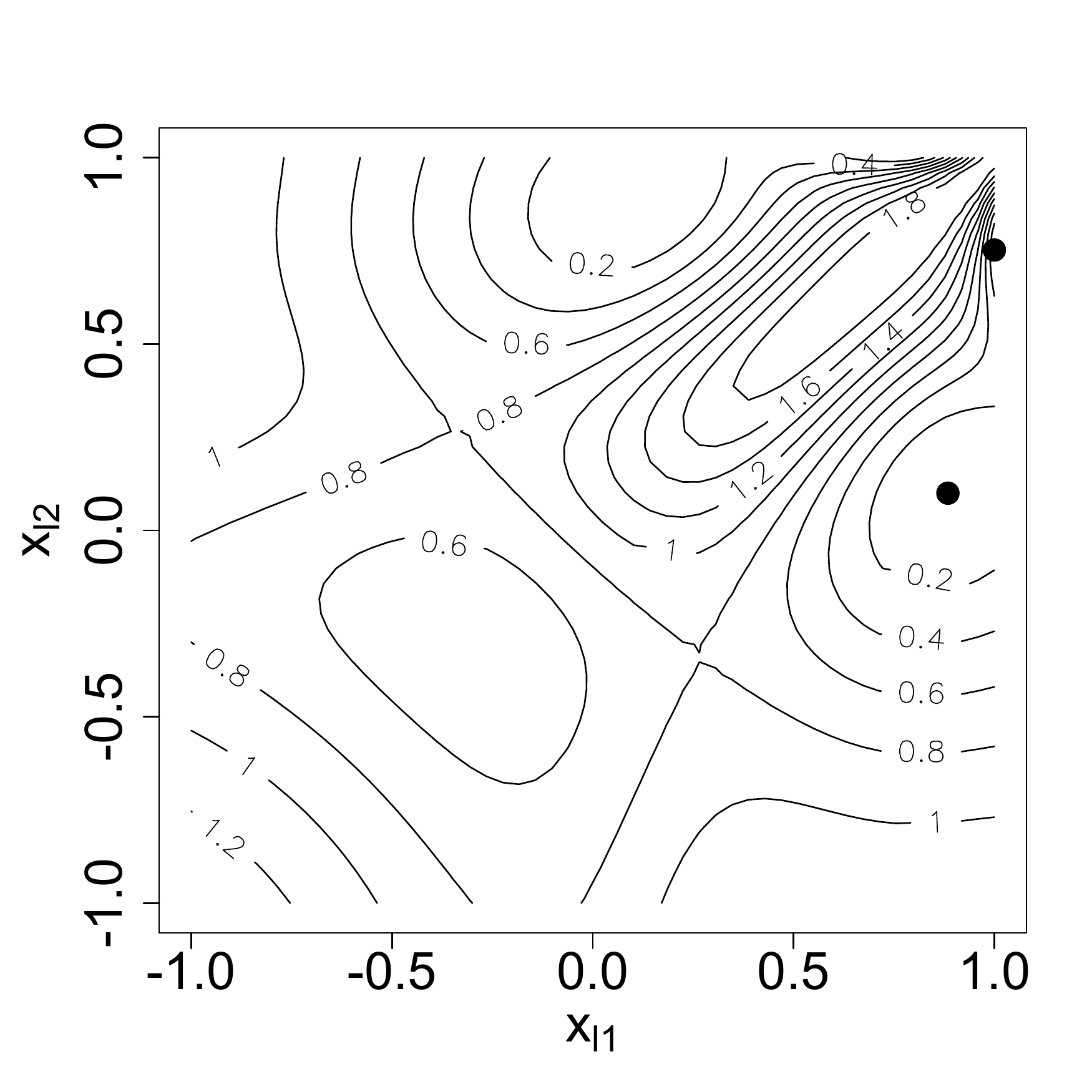} & \includegraphics[scale=0.375]{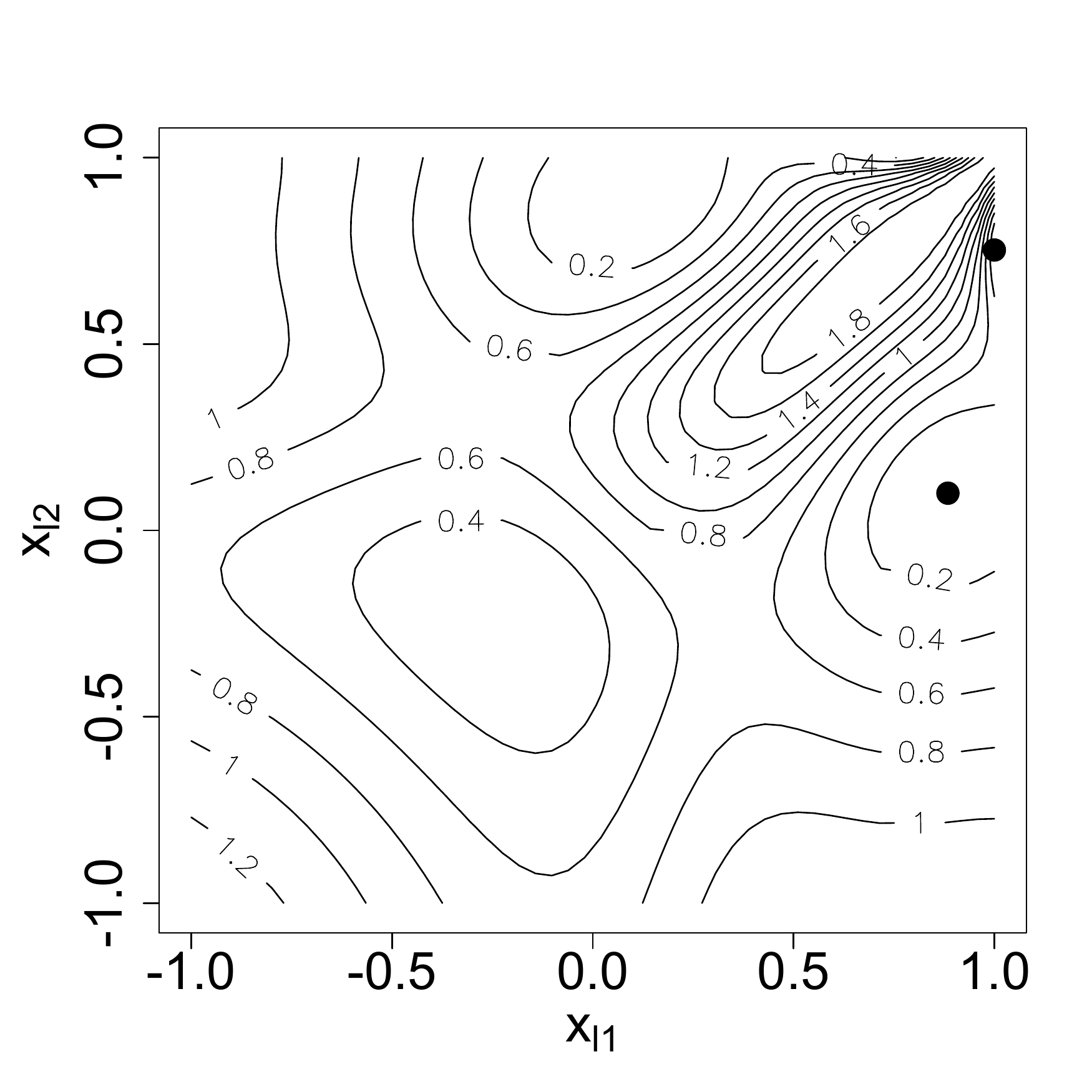} \\
\includegraphics[scale=0.375]{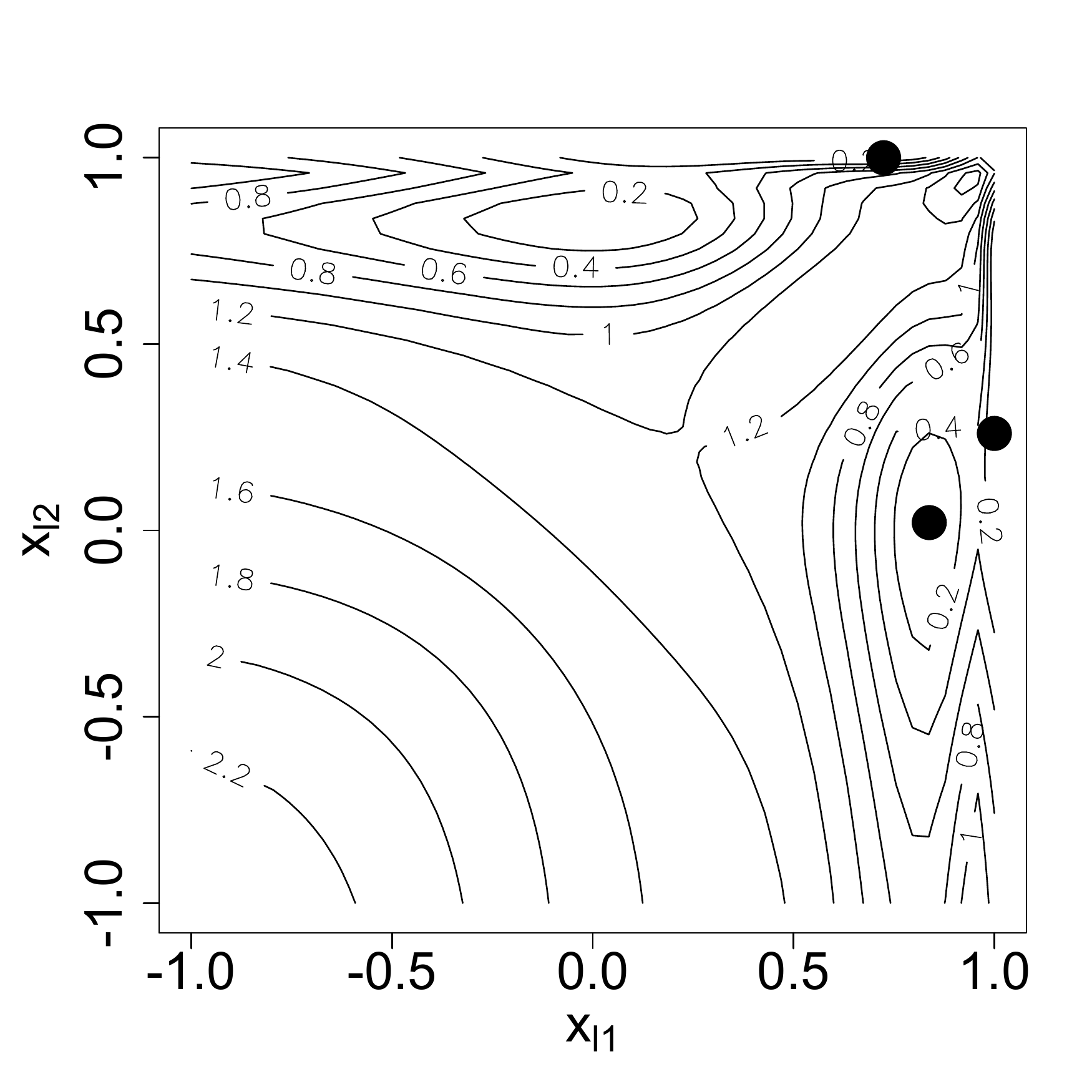}
\end{tabular}
\end{center}
\caption{\label{mvgetplot}Derivative surfaces, $\psi(\boldsymbol{x},\xi^\star;\,\boldsymbol{\tau})$, and D-optimal designs for QL (top left), MQL (top right) and GEE (bottom left) approximations.}
\end{figure}

Optimality of these designs can be confirmed, as in \S\ref{theorsec}, via the application of a multivariate equivalence theorem; see \citet{aca:2008b} and \citet{wv}. A necessary and sufficient condition for a design $\xi^\star$ to be locally D-optimal is

\begin{equation}\label{mvget}
\psi(\boldsymbol{x},\xi^\star;\,\boldsymbol{\tau}) = p-\mbox{trace}\left\{\boldsymbol{M}\left(\zeta\,;\boldsymbol{\tau}\right)\boldsymbol{M}^{-1}\left(\xi^\star\,;\boldsymbol{\tau}\right)\right\} \ge 0\,,
\end{equation}

\noindent for all $\zeta\in\mathcal{X}^m$. This condition can be verified numerically; Figure~\ref{mvgetplot} plots the derivative function for each of the three approximations, with the support points of the optimal designs marked. Notice that: (i) the support points occur at minima of the derivative surface, with $\psi(\boldsymbol{x},\xi^\star;\,\boldsymbol{\tau})=0$; (ii) with blocks of size $m=2$, the derivative function must be symmetric in the line $x_{l1}=x_{l2}$; and (iii) that the derivative surfaces for QL and MQL are very similar.

\section{Extensions and further reading}
\label{extensec}

Although we have described designs for three types of response and several link functions, interest in designs for generalized linear models continues to focus on binary responses and the logistic link. Much of this reflects the rapid growth of applications of the discrete choice models described in detail in Chapter 26.

There is also appreciable interest in design for logistic regression in medical statistics, particularly drug development. A typical problem is to find the dose giving a specified probability of toxicity. The natural way to proceed is the  use of a sequential design as described in \S\ref{seqsubsec}. The appropriate design criterion is c-optimality, which  designs to estimate this dose with minimum variance. Such designs for nonlinear regression models are exemplified by \citet[\S17.5]{ADT:2007} and by \citet{ftw:92} for GLMs. Often, however, the locally optimal design has a single support point  at the estimated dose. The sequential design may then fail to provide sufficient information to guarantee identification of the required dose \citep{luc:2000,squigley:2010}. The designs need to provide   sufficient perturbation in the experimental conditions to ensure convergence.

The designs we have exemplified, particularly for first-order models, often have the number of support points equal to the number of parameters. They therefore fail to provide any information for model checking and choice. Designs for discriminating between two regression models were introduced by \citet{fedorov1972optimal} and by \citet{a+fed:75a} who called them T-optimal. \citet{ponce+a:92} and \citet{Waterhouse2008132} extend T-optimality to GLMs. A  general design criterion for discrimination between models using Kullback-Leibler distances is that of \citet{jesus+:2007}.  A potential disadvantage of these designs is that they focus exclusively on model discrimination. Compound designs for the joint problem of parameter estimation and model discrimination, called DT-optimal, are given, for regression models, by \citet{aca:2008a}. \citet{Waterhouse2008132} also attend to the quality of parameter estimates, but not through use of a compound criterion. D-optimal designs robust to the form of the linear predictor were developed by \citet{woods+:2006} and \citet{wl2011}.

In some applications it is not necessary to establish a model that holds globally. In the context of dose finding, \citet{squigley:2010} recommends the use of a model that has adequate local behaviour. A more formal approach to model uncertainty is in Chapter 24. In particular, \citet{li+wiens:2011}, considers approximate models in dose response experiments, whereas \citet{wiens:2009} provides robust designs for discrimination between regression models.

In \S\ref{gamsec} we mentioned that response transformation is sometimes an alternative to the use of a gamma model. \citet{a+cook:96} find optimal designs for estimation of the transformation in linear models whereas \citet{aca:2005} studies  designs for transforming both sides of a nonlinear model.

Finally, we note that some interesting models for non-normal responses have a structure as in   \S\ref{famsec} but with a predictor which is nonlinear in the parameters. Optimal designs for one particular generalized nonlinear model are given by \citet{steff+dave:2011}.

\section*{Acknowledgments}
The authors thank Dr T.W. Waite (University of Southampton, UK) for assistance with the computation in \S\ref{corr}. D.C. Woods was supported by a Fellowship from the UK Engineering and Physical Sciences Research Council (EP/J018317/1), and the work was partly undertaken while the authors were Visiting Fellows at the Isaac Newton Institute for the Mathematical Sciences (Cambridge, UK).

\bibliographystyle{chicago}

\end{document}